\documentclass[useAMS,usenatbib]{emulateapj}
\usepackage{epsfig}
\usepackage{lineno}
\usepackage{epsfig} 
\usepackage{amsmath}
\usepackage{amssymb}
\usepackage{natbib}
\usepackage{graphicx}
\usepackage{endnotes}

\shorttitle{timining and spectral variability studies of IGR J17091$-$3624}
\shortauthors{Pahari et al.}

\begin{document}

\title{X-RAY SPECTRAL STATE EVOLUTION IN IGR J17091$-$3624 AND COMPARISON OF ITS HEARTBEAT OSCILLATION PROPERTIES WITH THOSE OF GRS 1915+105}
\author{Mayukh Pahari$^{1}$, J S Yadav$^1$ and Sudip Bhattacharyya$^1$}
\affil{$^1$ Tata Institute of Fundamental Research, Homi Bhabha Road, Mumbai, India; \texttt{mp@tifr.res.in\\}} 

\linenumbers

\begin{abstract}

In this work, we study the X-ray timing and spectral evolution of the transient low mass X-ray binary IGR J17091$-$3624 during first 66 days of its 2011 outburst. We present results obtained from observations with two instruments, {\it Rossi X-ray Timing Explorer (RXTE)} Proportional Counter Array (PCA) and {\it SWIFT} X-ray telescope (XRT), between 09 February, 2011 and 15 April, 2011. Using quasi-periodic oscillation classifications, power density spectrum characteristics, time-lag behavior and energy spectral properties, we determine source states and their transitions at different times of the outburst. During the first part of the evolution, the source followed trends usually observed from transient black hole X-ray binaries (BHXBs). 
Interestingly,  a gradual transition is observed in IGR J17091$-$3624 from the low-variability SIMS, commonly seen in BHXBs, to a high$-$variability state with regular, repeatitive and structured pulsations, seen only from GRS 1915+105 (also known as `$\rho$' class variability/`heartbeat' oscillations). We study the time evolution of characteristic time$-$scale, quality factor and rms amplitude of heartbeat oscillations in IGR J17091$-$3624. We also present a detailed comparison of the timing and spectral properties of heartbeat oscillations and their evolution in IGR J17091$-$3624 and GRS 1915+105.  

\end{abstract}

\keywords{accretion, accretion disks --- black hole physics --- X-rays: binaries --- X-rays: individual: (IGR J17091$-$3624, GRS 1915+105)}

\section{Introduction}\label{intro}

IGR J17091$-$3624 was detected as a bright, variable, transient source with flaring activity in October 1994 (Mir/KVANT/TTM), September 1996 (BeppoSAX/WFC), September 2001 (BeppoSAX/WFC; \citet{b88}), April 2003 (INTEGRAL/IBIS; \citet{b19}), and July 2007 \citep{b3}.
In April 2003, IBIS, the Gamma-ray imager on board the {\it INTEGRAL} satellite detected the transient X$-$ray binary IGR J17091$-$3624 at the position of 17$^h$09.1$^m$ (R.A. (J2000.0)), -36$^{\rm o}$24.63$^{'}$ (declination (J2000.0)) while monitoring the Galactic centre \citep{b19}. 
 A transition between soft and hard states and the presence of a blackbody component during source softening were observed during a joint spectral analysis of {\it INTEGRAL}/IBIS and {\it RXTE}/PCA data by \citet{b1}. They found that the different spectral states, their transitions, and time variability of the source were similar to those of the black hole candidate H1743$-$322. 
ek dustu haoa bhese aase
The approach which is commonly used to define the spectral states of a source during an outburst evolution is based on energy spectral properties as well as power density spectral (PDS) properties including low$-$frequency quasi-periodic oscillations (LF QPOs) features and classifications \citep{b83}. Depending on QPO frequency, coherence and strength, three main types of LF QPOs are observed in BHXBs -  type-A (weak (with few percent of rms amplitude) and broad (q-factor $\le$ 3) QPO around 7$-$9 Hz; \citep{b81,b30}), type-B (relatively strong ($\sim$4-7\% rms amplitude) and relatively narrow (q-factor $\sim$5$-$7) QPO around 5$-$7 Hz appeared with a weak red noise \citep{b81, b30}) and type-C (strong (up to $\sim$21\% rms amplitude) and narrow (q-factor $\sim$ 5$-$12) features with the variable centroid frequency between 0.1$-$15 Hz, superposed on a strong flat-top noise \citep{b81,b30}). Type-B QPOs are generally associated with relatively soft state or soft intermediate state (SIMS) \citep{b86, b83} while type-C QPOs carry the signature of the hard intermediate state (HIMS; \citep{b86, b83}). 
Along with different types of QPOs, one important difference, which has also been used in this work to distinguish the HIMS from the SIMS is that type-B QPOs always show hard lag (hard photons lag soft photons) at the fundamental QPO frequency between 5$-$7 Hz, but a type-C QPO, if detected in the frequency range 5$-$7 Hz during the HIMS, always shows soft lag at the fundamental QPO frequency \citep{b81}. The low/hard state (LS) is usually characterized by strong band-limited noise where noise component can be decomposed into a few broad Lorentzian components \citep{b86}. 
 
\citet{b22} found that the PDS of IGR J17091$-$3624 could be well described with the standard model of band-limited noise with the break frequency of 0.31 $\pm$ 0.04 Hz. They fitted the energy spectrum with the power-law model having the power$-$law index of 1.43 $\pm$ 0.03. These timing and spectral properties are commonly observed in the LS of BHXBs \citep{b91}. Using combined {\it RXTE}/PCA and {\it INTEGRAL}/IBIS data, \citet{b1} showed that a single thermal Comptonization component ({\tt compTT} in {\tt XSpec}) or saturated Comptonization ({\tt compST} in {\tt XSpec}) or the combination of soft multi-color disk blackbody and hard power-law ({\tt diskbb+powerlaw}) are good description of the X-ray spectra of IGR J17091$-$3624. However, they noticed that the model derived electron temperature was $\sim$ 20 keV, which was less than typical electron temperature of black hole X-ray binaries ($\sim$ 50-100 keV; \citet{b45}). 
{\it SWIFT}/XRT observations of IGR J17091$-$3624 during the outburst in 2007 revealed another state transition from hard to soft, and refined the X-ray position of this source, which is consistent with the position of the infra-red counterpart observed with the ESO {\it New Technology Telescope} (NTT) \citep{b24} as well as compact radio counterpart in the {\it VLA} archived data \citep{b3}. 
 
 The renewed activity of the source was first detected with the {\it SWIFT}/BAT when the hard X-ray flux (15.0-60.0 keV) increases up to 60 mCrab on 03 February, 2011 \citep{b11}. Later using the {\it SWIFT}/XRT observation, \citet{b10} confirmed that the outburst occurred in the source IGR J17091$-$3624, and not in IGR J17098$-$3628, which is also an X-ray transient 9.6$^{'}$ away from IGR J17091$-$3624. 
The {\it INTEGRAL}/IBIS also detects the source simultaneously with the {\it SWIFT}/XRT \citep{b8}. The preliminary spectral data during the recent outburst of IGR J17091$-$3624 in 2011 showed that a soft spectral component is necessary to describe the spectrum along with the hard power-law component \citep{b8}. Later using the data from IBIS/ISGRI imager and the JEM-X telescope on board the {\it INTEGRAL} satellite on 7$-$8 February, 2011, \citet{b25} described the combined ISGRI and JEMX2 spectrum (5.0-300.0 keV) with a power law having the high energy cut-off at 110$^{+47}_{-27}$ keV. In spite of several efforts, little is known about the compact nature of this source or the nature of its binary companion (but see \citet{b82}). 

From the beginning of {\it RXTE}/PCA observations on 9 February, 2011, several rich features in timing and spectral domain of the source have been noticed. QPOs at 94 $\pm$ 3 mHz and 0.105$^{+0.004}_{-0.003}$ Hz with a quality factor of $\sim$ 9.1 and $\sim$ 7.4 respectively, were reported using {\it RXTE}/PCA observations on 9, 10, 11 and 12 February, 2011 \citep{b13} although a source contamination problem exists with the RXTE data during this period.  Apart from this, fast QPO evolution \citep{b12}, 10 mHz QPOs \citep{b6} and evolving power density spectrum with multiple peaks \citep{b50} were also noticed. High frequency QPO at 66 Hz \citep{b94} was detected with the significance of 8.5$\sigma$.
 Interestingly, a quasi-regular oscillation in the light curve of IGR J17091$-$3624 is noticed which is very similar to the `$\rho$' class variability seen only in the light curve of the Galactic micro-quasar GRS 1915+105 \citep{b26}. Later, \citet{b95} claimed a few more variabilities similar to those found in GRS 1915+105. Two new variabilities, which are unique to IGR J17091$-$3624, were reported by \citet{b28}.  Using Chandra spectroscopy of IGR J17091$-$3624 during soft, disk-dominated state, \citet{b96} showed that accretion disk wind is anti-correlated with relativistic jets. This disk wind was observed to be highly ionized, dense, and had typical blue-shift of $\sim$9300 km/s, or less projected along our line of sight. Dips in the X-ray intensity, which are observed in the $\kappa$/$\lambda$ class of GRS 1915+105, were also found in IGR J17091$-$3624 with similar properties \citep{b61}. 

GRS 1915+105 is a Galactic micro-quasar which has shown spectacular X-ray variabilities since its discovery in 1992 \citep{b14}. Among different variability classes in GRS 1915+105 \citep{b16}, the `$\rho$' class shows regular, repetitive bursts in the light curve with the recurrence time between $\sim$31 sec and $\sim$106 sec. These recursive burst structures have been observed for extended duration and have been discussed by several authors \citep{b2,b40,b5}. Using the data from the Indian X-ray Astronomy Experiment (IXAE) on board the Indian Satellite {\it IRS-P3}, as well as {\it RXTE}, \citet{b5} showed that these bursts which could be categorized as irregular, quasi$-$regular and regular bursts have a slow exponential rise and sharp decay. Using flux-resolved spectroscopy, \citet{b62} showed that large, repeatitive oscillations in GRS 1915+105 appeared on the top of persistent emission, are consistent with the slim accretion disk approximations. Using the {\it BeppoSAX} data, \citet{b4} concluded that the process, responsible for the pulses, produces strongest emission between 3.0 and 10.0 keV, and the emission at the rising phase of the pulse dominates the lower and higher energies. However, this hypothesis is not tested. Hence the consensus on the resolution of the issue with the origin of `$\rho$' class activity in GRS 1915+105 is yet to achive. The similarity of the $\rho$ class of GRS 1915+105 with the variability observed in IGR J17091$-$3624 motivates us to perform a detailed analysis of the current outburst of IGR J17091$-$3624.

In the present work, we consider all observations of IGR J17091$-$3624 between 09 February, 2011 and 15 April, 2011 with the {\it RXTE}/PCA and the {\it SWIFT}/XRT. We have analyzed both timing and spectral properties of the source after confirming that the contamination from the nearby source IGR J17098$-$3628 is usually very small. In section 2 we describe our analysis methods. Section 3 summarizes results regarding the hard and soft intermediate state, a fast source evolution and the transition to a `$\rho$'-like class via an intermediate variability state, observed from IGR J17091$-$3624. Finally we compare the timing and spectral properties of the `$\rho$'-like class in IGR J17091$-$3624 with the `$\rho$' class characteristics seen from GRS 1915+105, and discuss their implications in section 4. We give our conclusions in section 5.                   

\section{Observations and Data Analysis}

We analyze all {\it RXTE}/PCA observations of IGR J17091$-$3624 between 9 February, 2011 (MJD 55601) and 15 April, 2011 (MJD 55666). Due to insufficient good time intervals and inferior quality, we are unable to extract the background data file for all High-Energy X-ray Timing Experiment (HEXTE; \citet{b106}) observations. Hence, we use only the {\it RXTE}/PCA {\tt Standard-2} Science Array data and event mode {\tt GoodXenon} data for our analysis. Observation details are given in Table 1. We also analyze five {\it RXTE}/PCA observations of GRS 1915+105 between 26 May, 1997 and 22 June, 1997 when the source was in `$\rho$' class. {\it RXTE}/PCA consists of an array of five  proportional counters (PCU0-4) filled with Xenon gas, with total effective area of $\sim$ 6500 cm$^{2}$, operating in 2.0-60.0 keV energy range \citep{b27}. We use data observed with all Xenon layers of {\tt PCU2}. During the start or the stop of any PCU unit, many observations show large count rate fluctuations in the light curve due to the instrumental effects. To take care of that, we first create xte filter file for each PCA observation and calculate good time interval (GTI) from the exposure time by applying all standard filtering criteria including breakdown events. Calculated values of good time interval are provided in Table 1. Using GTI values of individual observation, we extract background subtracted light curve from {\tt GoodXenon} data having resolution of 125 $\mu$s. Observations with the background subtracted PCA intensity $\ge$ 15 cts.s$^{-1}$PCU$^{-1}$ are considered. 

We also analyze 28 archival data sets of {\it SWIFT}/XRT between 03 February, 2011 and 30 March, 2011. The XRT instrument on board {\it SWIFT} satellite has the effective area is 110 cm$^{2}$ at 1.5 keV, the position accuracy is $\sim$ 5 arc sec and is operating in 0.2-10.0 keV energy range \citep{b32}. The typical spectral resolution of this instrument is $\sim$ 140 eV at 6.0 keV. To avoid pile-up problem while observing luminous X-ray sources, all observations (except the first one) are performed with Windowed Timing (WT) mode which has time resolution of 2.2 ms and flux limit of $\sim$ 600 mCrab ($\sim$ 200 cts/s) \citep{b33}. For our analysis, we use level 2 cleaned event data which has been extracted from level 1 calibrated data after applying screening criteria on specific parameters like CCD Temperature, Sun Angle, Elevation Angle etc. For all WT mode data, we assign grade 0-2 and selected good event with {\tt STATUS==b0}. We extract source and background light curves and spectra separately using {\tt xselect v2.4b} in {\tt FTOOLS 6.10}.

Top panel of Fig. 1 shows the background-subtracted average {\it RXTE}/PCA count rate between 2.0-60.0 keV, while middle panel shows the background-subtracted average {\it SWIFT}/XRT count rate between 0.3-10.0 keV. HIMS \& SIMS, intermediate variable state (IVS) and variable state are shown by open circles, stars and triangles respectively in the top and bottom panels of Fig. 1 while all {\it SWIFT}/XRT observations are shown by solid circles in the middle panel.
During first 14 days of observation, IGR J17091$-$3624 was within the field of view of persistently luminous Z-type neutron star low mass X-ray binary GX 349+2 (17$^h$05$^m$44.49$^s$ (R.A. (J2000.0)), -36$^{\rm o}$25.38$^{'}$ (declination (J2000.0))). As a consequence, very high PCA count rate was observed in first 14 days of observation compared to the rest. Therefore, in our present work, we consider PCA observations from 23 February, 2011 onwards (observations falls right to the first vertical dotted line in the top panel of Fig. 1) to avoid the contamination effect. Further to check possible contamination from IGR J17098-3628, a transient BHXB located 9.6$^{'}$ away from IGR J17091$-$3624, we analyse background-subtracted image from {\it SWIFT}/XRT PC mode data in the energy range 0.3-10.0 keV (Fig. 2). With {\it SWIFT}/XRT, no X-ray source has been detected at the position of IGR J17098-3628 (shown by a circle), but is clearly detected at the position of IGR J17091$-$3624 (shown by a square). This is consistent with earlier results \citep{b10}. For a quantitative measurement, we calculate the count rate separately from the position of IGR J17091$-$3624 and from the position of IGR J17098-3628. For the energy range 0.3-10.0 keV, their values are 2.023 cts/s and 0.213 cts/s respectively on 3 February, 2011. So, at maximum, the intensity of IGR J17098-3628 is $\sim$ 10\% of IGR J17091$-$3624. Since, {\it RXTE}/PCA could not spatially resolve these sources, it is expected that at maximum $\sim$ 10\% contamination effect would be present in {\it RXTE}/PCA data of IGR J17091$-$3624. 

\subsection{Timing Analysis}

For each PCA observation we create rms normalized and white noise subtracted power density spectrum (PDS) to study variability features and to track the spectral evolution of the source state with time. The dead time corrections are applied to the light curves. For {\tt GoodXenon} data, the dead time per event is approximately 10 $\mu$s. For details of the procedure, please check RXTE cookbook\footnote[1]{http://heasarc.nasa.gov/docs/xte/recipes/pca\_deadtime.html}. To improve the signal-to-noise ratio in the PDS, we use the geometric re-binning of frequency bins by the factor of 1.02\footnote[2]{http://heasarc.nasa.gov/xanadu/xronos/help/powspec.html}. The expected white noise is subtracted from the PDS$^2$ and they are normalized such that their square root of integral over the range of frequencies provide fractional rms variability. We produce PDS from the {\it SWIFT}/XRT data using background subtracted light-curve between 0.2-10.0 keV energy range. Since one can not compare the 2.0-60.0 keV PDS from the PCA data with 0.2-10.0 keV PDS from {\it SWIFT}/XRT data, hence we extract only {\it SWIFT}/XRT PDS for observation dates when no {\it RXTE}/PCA data are available. Different types of QPOs and noise components are noticed in the PDS of different {\it RXTE}/PCA observations. The details of PDS features are given in Table 1. 

\subsection{Spectral Evolution}

The color-color diagram (CD), the hard color vs. soft color plot and the hardness intensity diagram (HID) are created using the PCA {\tt GoodXenon} data with 1.0 s bin time. The soft color is computed by dividing the background subtracted PCA count rates in the energy band 5.0-12.0 keV by that in the energy band 2.0-5.0 keV. Similarly, the hard color is defined as the ratio of the background subtracted count rate in the energy band 12.0-60.0 keV and 2.0-5.0 keV. X-ray intensity is defined as the background subtracted PCA count rate in 2.0$-$60.0 keV energy range. Since the channel gain varies from epoch to epoch, we use absolute channel ranges corresponding to each of the energy ranges to calculate soft color, hard color and X-ray intensity for the present epoch. For each {\it RXTE}/PCA observation, mean count rate (top panel of Fig. 1), and mean value of hard color and soft color are provided in Table 1 with errors.

\subsection{Spectral Fitting}\label{SpectralAnalysis}

In case of {\it RXTE}/PCA, we perform the spectral analysis using the {\tt Standard-2} data. While fitting, we add 1\% systematic errors. In case of {\it SWIFT}/XRT data, we first extract an 1D image of the source. Then choosing suitable source region and background region, we extract source spectrum and background spectrum separately from the cleaned and pointed WT mode event file. We use the latest standard response matrix file (rmf) and create auxiliary response file (arf) using exposure map for the WT mode data between 0-2 grade values. We use {\tt XSpec v 12.6.0} for spectral fitting of {\it RXTE}/PCA and {\it SWIFT}/XRT data. We use 3.0$-$25.0 keV energy range for {\it RXTE}/PCA spectral analysis since the source count rate statistics is poor outside this range. We use {\it SWIFT}/XRT spectra in the energy range of 0.6$-$10.0 keV. 

We have tried different single component models like {\tt diskbb}, {\tt compTT} or {\tt powerlaw} as well as dual component models like {\tt diskbb+powerlaw}, {\tt diskbb+compTT} or {\tt compTT+powerlaw} to fit the source spectra. The reason for selecting and testing different models is discussed in \S~\ref{intro}. We find that the single {\it SWIFT}/XRT spectrum could be fitted well with a combination of {\tt diskbb} and/or {\tt powerlaw} model. Fitting with the rest of the model combination yields unacceptably large reduced $\chi^2$ ($>$ 1.5) or sometimes over-fit the spectra due to large no. of parameters. This is also true for {\it RXTE}/PCA spectra. Using INTEGRAL energy spectra, high energy cut-off ($>$ 30 keV) has been detected in this source \citep{b92}. A broadband  BHXB spectrum, during a SIMS or HIMS, should be well described also by a combination of disk and Comptonization. Since we restrict our {\it RXTE}/PCA spectral analysis to 25 keV, we are unable to detect cut-off or Comptonization where electron temperature is pretty high. Thus, a relatively simple model can describe our spectra well. It may be noted that \citet{b93} showed that {\it RXTE}/PCA spectra in transient BHXBs can be well described with a multi-temperature disk blackbody emission and a hard power-law emission with/without cut$-$off. While fitting {\it RXTE}/PCA spectra, we use a small Gaussian component at 6.4 keV in order to improve the reduced $\chi^2$. 
To account the effect of absorption by the neutral Hydrogen, all model components are multiplied by a photo-electric absorption model. The Galactic absorption column density at the direction of 17$^h$11$^m$12.5$^s$ (R.A. (J2000.0)), -36$^{\rm o}$24.6$^{'}$ (declination (J2000.0)) is calculated as 0.79 $\times$ 10$^{22}$ cm$^{-2}$. From spectral analysis, we find that for all observations, absorption column density varies between 0.82 $\pm$ 0.2 and 0.99 $\pm$ 0.3. Thus we keep this value fixed at 0.9 $\times$ 10$^{22}$ cm$^{-2}$ for all spectral fitting. 

\section{Results}

{\it SWIFT}/XRT observations of IGR J17091$-$3624 show a transition of the source from the low count rate variability to the high count rate variability (see lower top panel of Fig. 1). Later, a gradual transition in the variability, i.e., from an irregular, less-variable state, to a regular, repetitive and large variability state has been observed. The large variability looks very similar to the unique variability observed during the `$\rho$' class in GRS 1915+105 \citep{b26}. This motivates us to perform a detailed study on the source spectral state evolution and compare it with that of GRS 1915+105. To distinguish among variability classes, we perform the count rate fractional rms analysis of different observations (by dividing the square root of the variance with the mean count rate) from {\it RXTE}/PCA which is shown in the bottom panel of Fig. 1. Later, comparing with timing and spectral properties, we show that observations which have fractional rms values less than 5 are consistent with the HIMS and SIMS, and observations with fractional rms significantly higher than 5 are consistent with the variable state/$\rho$-like class. A few observations show the transition between soft intermediate state and variable state/$\rho$-like class, and hence described as `Intermediate variable state (IVS)'. The details are discussed below.

\subsection{Low/hard state, hard intermediate and soft intermediate state}

From 14 February, 2011 to 24 February, 2011, a gradual rise in the {\it SWIFT}/XRT flux between 0.3-10.0 keV is observed in IGR J17091$-$3624 (see middle panel of Fig. 1). Till 14 February, 2011, average {\it SWIFT}/XRT count rate was $\sim$ 10 cts/s. Then it gradually increases to $\sim$ 45 cts/s on 24 February, 2011. To check whether this rise in flux led to any state transition, we perform a detailed study on timing properties of the source using {\it SWIFT}/XRT data. From 09 February, 2011 to 18 February, 2011 the PDS usually show band-limited noise, sometimes with the rms amplitude as strong as $\sim$ 9\% (see Table 1) and can be decomposed into multiple broad Lorentzians. Apart from this, a break in the PDS is observed once at $\sim$0.19 Hz. Broad but strong QPO-like features are observed at least twice at $\sim$ 0.2 Hz (with Q-factor $>$ 2; see Table 1). All these features in the PDS can easily be associated with the low/hard state (LS) \citep{b86}. On 20 February, 2011, a type-C QPO at $\sim$0.51 Hz along with the strong flat-top noise ($\sim$ 7\% rms) is observed in the PDS, which carries the signature of the hard intermediate state (HIMS). This imply that the source undergoes state transition from the LS to the HIMS. On 22 February, 2011, a broad Lorentzian (quality factor $<$ 2) is observed at $\sim$ 1.8 Hz along with weak red noise ($\sim$2-3\% rms) in the {\it SWIFT}/XRT spectra which may indicate that the source undergoes another transition from the HIMS to the soft intermediate state (SIMS). Detection of the type-B QPO on 23 February, 2011 using {\it RXTE}/PCA (see Table 1) confirms the transition \citep{b86}. \citet{b92} also observed the LS/HIMS to SIMS transition based on the detection of optically thin radio emission on MJD 55623.57 (03 March, 2011). From 24 February, 2011 to 13 March, 2011, the source usually shows back and forth transition between the HIMS and the SIMS with the usual detection of type-C and type-B QPOs respectively (see Table 1). In Fig. 3, we have shown example of the HIMS observed on 26 February, 2011 and the SIMS observed on 03 March, 2011. Top panels show the PDS where the HIMS shows strong flat-top noise ($\sim$ 8\% rms amplitude) with a strong type-C QPO at 3.73 $\pm$ 0.09 Hz but the SIMS show weak red noise ($\sim$ 3\% rms amplitude) with a transient type-B QPO at 4.85 $\pm$ 0.08 Hz. For more accurate determination of the state, we perform time-lag analysis between the the soft energy band (1.8$-$4.2 keV) and the hard energy band (5.0$-$13.0 keV) using {\tt GHATS v1.0.2} (T. Belloni, private communications) where the lag formulation is based on cross-spectra technique. At the fundamental QPO frequency, HIMS shows the soft-lag (soft photons lag hard photons) of $\sim$ 8 msec (middle left panel of Fig. 3) while the SIMS at the fundamental QPO frequency shows hard-lag of $\sim$ 22 msec (middle right panel of Fig. 3). In both panels, lags at the fundamental QPO frequency are shown by dotted vertical lines. These observations are in fine tune with the characteristics of these states.  

We also carry out energy spectral analysis of observations between 09 February, 2011 and 25 March, 2011 respectively. From the fitted parameters in Table 2, we find that disk component is not significant while fitting the spectra observed between 09 February, 2011 and 20 February, 2011. A single power-law can adequately describe these spectra. From 20 February, 2011 onwards a soft disk component is essentially present in all spectral fitting and the power-law becomes steeper (see Table 2). This indicate that a hard-to-soft state transition takes place. This is consistent with the PDS characteristics where the transition from the LS/HIMS to the SIMS is observed. An example of {\it RXTE}/PCA spectral fitting of the HIMS and the SIMS are shown in the bottom left and bottom right panel of Fig.3 respectively. From Table 2, it is clear that all low flux state observations (till 19 February, 2011) have photon index value less than 1.8 and no disk component is present at the detectable limit of {\it SWIFT}/XRT. These along with the detection of band-limited noise meet the criteria of a low/hard state as observed in transient BHXBs \citep{b90,b93,b86}.  However, a single power-law still provides a good fit (an F-test between {\tt powerlaw} and {\tt diskbb+powerlaw} models yields a probability of 1.7 x 10$^{-10}$) of the energy spectra on 20 February, 2011, where the LS to the HIMS transition is observed. Insignificant disk contribution during this transition is may be either due to very low disk flux of the source or very low temperature of the disk ($<$ 0.6 keV). Disk flux contribution is $\sim$10-20\% of the total flux in other HIMS where disk is detected significantly (see Table 2). 

It may be noted that the X-ray spectral state transition is observed between 18-22 February, 2011 is based only on {\it SWIFT}/XRT data in the energy range 0.2-10.0 keV since there are no simultaneous observation of {\it RXTE}/PCA between 9-22 February, 2011. Hence the robustness of the result, based on XRT PDS, cannot be established confidently although it is consistent with the detection of radio flares on MJD 55601.3 and 55605.6 \citep{b92} and show correct trend in spectral evolution, usually observed from normal transients \citep{b90, b86}. Due to this ambiguity, we put question marks in first five observations of state classification column in Table 1, where PDS features are not clear. However, our spectral analysis agrees well with spectral parameters obtained from the joint spectral analysis of {\it INTEGRAL}/IBIS and {\it SWIFT}/XRT data by \citet{b98}. The relatively small difference between both results may be due to the (1) use of cut$-$off power-law (where cut$-$off energy $>$70 keV, see Table 1 of \citet{b98}) (2) different absorption column density (1.1 $\pm$ 0.3). 

The outburst evolution of the source through the LS, HIMS \& SIMS and frequent to and fro transition between the HIMS and the SIMS (see Table 1) indicate that the disk is unstable and the evolution pattern roughly follows the outburst evolution observed in normal transient BHXBs \citep{b90, b86}. 
  
\subsection{Intermediate variable state}

 During the continuous monitoring of the PCA light curve in IGR J17091$-$3624, we find a gradual transition in the count rate variability pattern from 12 March, 2011 to 19 March, 2011 in the energy range 2.0-60.0 keV. On 12 March, 2011, the source was in the SIMS when the light curve is less variable and very similar to the typical light curves seen in canonical BHXBs (see the top left panel of Fig. 4). On 19 March, 2011, the light curve shows a very regular repetitive pattern, high variability (peak count rate $\sim$ 3$-$4 times higher than the count rate at the persistent level; see lower-middle left panel of Fig. 4), and high count rate fractional rms (see bottom panel of Fig.1). We denote it by variable state/$\rho$-like class. Observations of average count rate that belong to the HIMS/SIMS and variable state/$\rho$-like class are separated by a dashed vertical line in Fig. 1. 

The transition between the SIMS and the variable state occurs through some semi-oscillatory intermediate stage on 14 March, 2011. The random noisy pattern which is observed during the SIMS on 12 March, 2011, appears to be superimposed on an X-ray intensity oscillation with the timescale of $\sim$80$-$100 sec in the light curve on 14 March, 2011 (upper-middle left panel of Fig.4). The superimposed pulsations on random fluctuations are called intermediate variable state (IVS). Although these intermediate phases are transient but it has been observed few times like on 18, 23, 24 \& 26 March, 2011, when the source oscillates between the SIMS like random fluctuations and the highly variable $\rho$-like pulsations (see Table 1). The rms normalized PDS during the IVS show characteristic oscillations in the frequency range of $\sim$ 11$-$14 mHz with a variable quality factor between 3$-$9. They are usually associated with very weak red noise having rms amplitude 2-4\% and weak, narrow QPOs (see Table 1 and upper-middle right panel of Fig. 4). The IVS on 24 March, 2011 shows broad Lorentzian noise along with a type-A QPO at $\sim$ 8 Hz. Recently, a 11 mHz oscillation is discovered in the PDS using {\it Chandra} data during the HIMS in the BHXB H1743-322 \citep{b54}. However, the origin of 11 mHz oscillations in both sources may be different. From the middle right panels of Fig. 4, it is clear that IVS is distinctively different from the variable state/$\rho$-like class because (1) mHz pulsations in the IVS never show harmonics which, on the other hand, is frequently observed in variable state (lower middle right panel of Fig. 4). The absence of harmonics in the IVS may be due to very low count rate statistics associated with the PDS. (2) The characteristic oscillation frequency of the variable state is always higher than that of IVS and (3) rms amplitude of the continuum in variable state ($\sim$ 10-15\%) is significantly higher than that of the IVS ($\sim$5$-$7 \%). The term `intermediate' is used as it is observed between the SIMS, where no mHz QPOs are observed (top right panel of Fig. 4), and the $\rho$-like class where very strong mHz oscillations are noticed. Thus, IVS can be treated as the beginning of the $\rho$-like class in IGR J17091$-$3624. To trace any change in spectral state during this transition, we plot the HID (bottom left panel of Fig. 4) and the CD (bottom right panel of Fig. 4). From the CD and the HID it may be noted that both IVS and SIMS nearly coincide with each other (shown in red and black) while the variable state shows a softer and brighter part (shown in blue) which is absent in both SIMS and IVS. Hence, IVS has similar characteristics of SIMS with large amplitude oscillations superimposed to random fluctuations. The energy spectral analysis of the IVS between 3.0$-$25.0 keV on 14 March, 2011 shows that the spectrum can be well fitted by the {\tt diskbb+powerlaw} model which shows parameters similar to the SIMS. Using numerical solutions, \citet{b55} noticed that the limit cycle oscillations are observed when the viscosity fluctuation amplitude are small in an unstable disk. The pulsating behavior disappears and random fluctuations are restored when stochastic fluctuation amplitude exceeds a critical limit. However, random fluctuations and mHz oscillations can also co-exist. This model can explain results qualitatively from Fig. 4.

\subsection{Variable state/$\rho$-like class}

The new variable state/$\rho$-like class remains stable for almost 25 days (19 March, 2011 $-$ 12 April, 2011; see Table 1) except 23, 24, 25 \& 26 March, 2011. The recurrent time scale of this variability is found to change with time. Fig. 5 shows the evolution of the source via five {\it RXTE}/PCA light curves from 19 March, 2011 to 10 April, 2011. On 19 March, 2011, the variability timescale of recursive bursts is $\sim$ 33.8 sec and the ratio of peak count rate to the persistent level count rate is 2.92 $\pm$ 0.05. These bursts have simple profiles with slow-rise and fast-decay. Light curves in Fig. 5 show that, as these bursts evolve with time, flaring frequency roughly increases. On 10 April, 2011, the recurrence time scale of the variability becomes $\sim$ 21.7 sec. and the burst becomes more structured. The ratio of peak count rate to the persistent level count rate becomes 3.38$\pm$0.06. Careful observation of burst profiles in Fig. 5 reveals two unique burst structures (1) Burst profile seen on 31 March, 2011 have a single peak and the count rate rises slowly ($\sim$ 30 sec) to the peak of the burst. (2) Burst profile seen on 03 April, 2011 can be characterized by double peak and show faster rise in count rate ($\sim$ 15-20 sec). This behavior and the variation in burst profile structures of the source has so far been seen in the `$\rho$' class of GRS 1915+105. From Table 1, it is clear that $\rho$-like variabilities in IGR J17091$-$3624 becomes stronger (i.e., rms amplitude of characteristic oscillations increases from $\sim$ 6\% to $\sim$ 32\% ) and faster (i.e., recurrence timescale decreases from $\sim$ 80 sec (on 24 March, 2011) to $\sim$ 22 sec (12 April, 2011)) with time. 

For each variable state observation, we fit the PDS continuum with the broken power$-$law model (since low frquency break is prominent; see top panel of Fig. 6) and we use Lorentzian to fit characteristic pulsation peaks and broad noise components. From Table 1, PDS fit shows that most of the time, the characteristic pulsation frequency appears along with the first harmonic (e.g., $\sim$ 30 and 62 mHz on 19 March,  $\sim$ 24 and 50 mHz on 29 March etc.), a few times with both first and second harmonics (e.g., $\sim$ 31, 63 and 110.4 mHz on 20 March, $\sim$ 34, 68 and 110.2 mHz on 02 April etc.). We also detect type-A QPOs around 7$-$9 Hz during this state. An example of fitted PDS during variable state on 31 March, 2011 is shown on the top left panel of Fig. 6. In the bottom left panel of Fig. 6, we show the time evolution of characteristic oscillation frequencies of $\rho$ class in both IGR J17091$-$3624 and GRS 1915+105. This panel indicates that IGR J17091$-$3624 shows faster time evolution of characteristic pulsation frequency than that of GRS 1915+105 by a factor of $\sim$ 1.3 (measured by the change in characteristic oscillation frequencies in the given period provided in the bottom left panel of Fig. 6).In order to check how these recurrent bursts loose their quality factors with time, we plot the q-factor and fractional rms amplitude (\%) of the variable state characteristic oscillations with time in the top right panel of Fig. 6. It shows a random fluctuation in the q-factor, which is, unrelated with the fractional rms amplitude as well as the  pulsation period. A closer inspection show that from day 20 onwards, the quality factor of characteristic pulsations overall drops by a factor of $\sim$ 6, while during the same time, fractional rms amplitude roughly increases. Thus, highly coherent characteristic pulsations in the variable state/$\rho$-like class seem to be changed into the broad Lorentzian component at the end of their evolution. This is also reflected from the PDS fit on 15 April, 2011 (see Table 1) when multiple broad Lorentzian components appear during the transition from the $\rho$-like class to some other variability classes (may be similar to the $\kappa$/$\lambda$ class in GRS 1915+105).  A correlation is observed between the pulsation frequency of characteristic oscillations and its fractional rms amplitude (\%) (shown in bottom right panel of Fig. 6) with the Pearson product moment correlation coefficient (PMCC) of 0.91 \citep{b56}. 

In the following section, we explore other properties of the variable state/$\rho$-like class and compare them with those of GRS 1915+105.
 
\subsubsection {Properties of variable state/$\rho$-like class in comparison to GRS 1915+105}

As the variable state/$\rho$-like class in IGR J17091$-$3624 looks similar to the `$\rho$' class of GRS 1915+105, we make a comparative study between them. 

All panels of Fig. 7 show the 2.0-60.0 keV light curve in both sources. To track the change in variability pattern, we consider two observations of variable state/$\rho$-like class in IGR J17091$-$3624 which show transition from the slow variability \& low peak flux  (31 March, 11; top left panel of Fig. 7) to the fast variability \& high peak flux (10 April, 11; bottom left panel of Fig. 7). Similar transition from the slow variability \& low peak flux on 26 May, 1997 (top right panel of Fig. 6) to the fast variability \& high peak flux on 22 June, 1997 (bottom right panel of Fig. 7) is also observed in GRS 1915+105. However, it is interesting to observe that the average peak count of bursts in GRS 1915+105 increases significantly while transiting from the slow variability to the fast variability ($\sim$ 4000 cts/s/PCU to $\sim$ 5000 cts/s/PCU) while it is increased slightly ($\sim$ 400 cts/s/PCU to $\sim$ 430 cts/s/PCU) in IGR J17091$-$3624 during the variable state/$\rho$-like class. It may be noted that the variability in IGR J17091$-$3624, although repetitive, is irregular, less structured (as the source is fainter than GRS 1915+105) and less coherent (i.e., q-factor is lowered by a factor of 1.4$-$1.8) compared to the `$\rho$' class of GRS 1915+105. In both sources, double-peak bursts are more frequent during fast variability regime.

Previously, a strong anti-correlation between the X-ray intensity and the hardness ratio (defined as the ratio of count rate between 12.0-60.0 keV and 2.0-12.0 keV) is found in the `$\rho$' class of GRS 1915+105 \citep{b5}. This anti-correlation is also found in the variable state/$\rho$-like class of IGR J17091$-$3624. Fig. 8 shows the comparative result on the anti-correlation found in IGR J17091$-$3624 on 31 March, 2011 (top and bottom left panels) and in GRS 1915+105 on 22 June, 1997 (top and bottom right panels). It is clear that at every second on burst profile, the hardness ratio is higher in IGR J17091$-$3624 than that observed in GRS 1915+105. 

To study the nature of bursts in details, we study the rise and decay profiles from both sources. Top left panel of Fig. 9 shows the combined rise profiles of several bursts and combined decay profiles of same bursts in IGR J17091$-$3624 on 31 March 11. To study the average behaviour of rise profiles, we normalize the starting time of each profile to 0 sec. Similarly, to study the average behavior of decay profiles, we normalize the starting time of the decay of each profile to 15 sec. We repeat the same procedure for the $\rho$ class observation in GRS 1915+105 on 22 June, 1997 (shown in the top right panel of Fig. 9) except for the fact that the staring time of each decay profile is normalized to 70 sec. We fit the combined rise profile with exponential rise function f$_{rise}$(t) = ae$^{t/t_{rise}}$ (red continuous line), the combined decay profile with exponential decay function f$_{decay}$(t) = ae$^{-t/t_{fall}}$ (blue continuous line) and all profiles with straight line (red and blue dotted line). In case of GRS 1915+105, exponential function fits better than straight line in both rise and decay profile with the significance of $>$ 10$\sigma$ and $>$8$\sigma$ respectively, where the significance is estimated using F-test. For IGR J17091$-$3624, the corresponding significance values are 3.2$\sigma$ and 5.1$\sigma$ respectively. The slope of the exponential function, for both rise and decay, are found similar in both sources (see Table 3). If burst structures depend on physical processes associated with the origin of the bursts \citep{b51}, then our results indicate that the origin of burst structure may be similar in both sources. We also study energy-dependent light curve variance spectra in both sources during their `heartbeat' oscillations. The variance spectrum in IGR J17091$-$3624 on 31 March, 2011 (bottom left panel of Fig. 9) shows that lightcurve variance decreases monotonously with energy. However, the variance spectrum in GRS 1915+105 on 22 June, 1997 (bottom right panel of Fig. 9) shows an initial increase in variance with the energy till $\sim$ 6 keV, then it decreases significantly, at least, up to $\sim$14 keV. This is remarkable as it may indicate that although similar parameter may trigger bursts in both sources, the spectroscopic evolution of the bursts is different, which, however, need to explore further.

Top panels of Fig. 10 show that the low frequency (0.1$-$10.0 Hz) continuum power, characteristic pulsation \& its harmonics and the nature of low frequency noise components are similar during $\rho$ class in both sources. White noise subtraction is not applied only for PDS shown in the top panels of Fig. 10. High frequency power in noise continuum becomes very weak above 1 Hz in case of IGR J17091$-$3624 (top left panel of Fig. 10 and top right panel of Fig. 4) although type-A/type-B QPOs are observed few times in the PDS. On the other hand, the strong, powerlaw like noise continuum in GRS 1915+105 continues at higher frequencies (at least up to 10 Hz; top right panel of Fig. 10) along with a type-B QPO at 6.7$\pm$0.4 Hz (middle right panel of Fig. 10).
Interestingly, the 6.7$\pm$0.4 Hz type-B QPO in GRS 1915+105 also has the harmonic at 13.2$\pm$0.8 Hz with the quality-factor of 3.8$\pm$0.4 (middle right panel of Fig. 10). Along with characteristic heartbeat oscillations, IGR J17091$-$3624 show 7$-$10 Hz type-A QPOs on 24, 25, 31 March, 2011 and 03 April, 2011 (see Table 1) and type-B QPO once at $\sim$ 5 Hz on 12 April, 2011 (see Table 1). May be because of statistical limit, harmonics in type-A/type-B QPOs are not visible in IGR J17091$-$3624. An example of type-A QPO observed during the $\rho$-like class on 31 March, 2011 at $\sim$ 8.4 Hz, is shown in the middle left panel of Fig. 10. Harmonics at mHz frequency are seen in both sources. Lower panels of Fig. 10 show the comparative study of the HIDs between two sources. We find similar trend in the HID evolution but different range in hard color for both sources. Results on comparative study of both sources are summarized in Table 3.

During 2011 outburst, IGR J17091$-$3624 evolves from the LS/HIMS to the SIMS state and finally from the SIMS to the variable state/$\rho$-like class (similar to $\rho$ class in GRS 1915+105) as X-ray flux increases (see Table 1 of \citet{b98} and Table 1 \& 2 from our analysis). Detection of accretion disk wind as well as quenching of radio emission during disk dominated state may suggest that IGR J17091$-$3624 is approaching the high soft state \citep{b96,b92}. 

According to the X-ray observations, IGR J17091$-$3624 passes through the SIMS state during radio flares observed on MJD 55623.57 \citep{b92}. Generally, HIMS to SIMS transitions are followed by Radio flares \citep{b86}. Following the SIMS, large amplitude X-ray oscillations appear via the intermediate variable state and continues for $\sim$ 25 days (MJD 55638 $-$ MJD 55663). Similarly, GRS 1915+105 was in the HIMS state during 26 March, 1997 $-$ 8 May, 1997 \citep{b99} and the $\rho$ X-ray class activity started on 26 May, 1997 and continue till 25 June, 1997 \citep{b5}. Between these periods (i.e., 8-26 May, 1997), for a short time, GRS 1915+105 showed Radio flare on 15 May, 1997 \citep{b97}. We analyse and fit the {\it RXTE}/PCA PDS on 13 May, 1997 using the combination of broken powerlaw and Lorentzian and find a type-B QPO at 5.87$\pm$0.12 Hz, with the quality factor of 7.89$\pm$0.62 and fractional rms amplitude (\%) of 4.84$\pm$0.36. The PDS is associated with weak red noise. These two observations show that the source was in the SIMS state on 13 May, 1997. Thus, GRS 1915+105 also passes through the SIMS before its large amplitude oscillations begin. This indicates that the X-ray spectral evolution prior to the $\rho$ class activity in both sources are roughly similar. This also strengthen that the spectroscopic nature of the origin of $\rho$ class activities in both sources may be similar. 

\section{Discussion}\label{discussion}

In this paper, we study the evolution of the X-ray activity in IGR J17091$-$3624 from 3 February, 2011 to 15 April, 2011. Initially, with the increase in the {\it SWIFT}/XRT count rate, transition from the low/hard and hard intermediate state to the soft intermediate state has been detected. In transient BHXBs,  this transition is usually accompanied by Radio flares \citep{b83, b86}. Our results show that the source makes to and fro transition between the HIMS and the SIMS a few times. This may represent the disk instability which occurs due to the increasing mass accretion rate. A few days later, a transition takes place from the SIMS to a regular, repetitive and highly variable pulsations similar to the $\rho$-class variability, previously observed only from GRS 1915+105. A few observations between variable state/$\rho$-like class and the SIMS show $\sim$ 11-14 mHz quasi$-$periodic variabilities with low rms amplitude which we term as intermediate variable state.  

In addition to the structural similarities of `heartbeat' oscillations in both sources, we find that they show similar time evolution in the burst structure. With time, characteristic pulsation frequency increases in both sources as well as the ratio of the peak flux to the persistent level flux also increases. In both cases, average rise profile of the burst shows slow exponential rise while average decay profile shows fast decay. Interestingly, we find strong anti-correlation between hardness value and the X-ray flux during variabilities in both sources. The evolution of the burst structure in the HID of both sources is identical. PDS below 1 Hz, show similar noise continuity as well pulsation peaks. Both sources have strong mHz QPOs along with harmonics and strong flat-top noise along with the presence of type-A/type-B QPOs in the PDS.

In other BHXBs, transition from the canonical low/hard state to the steep power law (SPL) state (highest flux regime of the outburst) \citep{b93}, which sometimes coincides with the soft intermediate state \citep{b83, b86}, is a commonly observed phenomenon. IGR J17091$-$3624 also shows transition from the LS/HIMS to the SIMS and the high soft state. Spectral properties of the low/hard state, HIMS and SIMS in IGR J17091$-$3624 also match with those of other BHXBs \citep{b93,b86}. \citet{b22} showed evidences that previously IGR J17091$-$3624 was in the low/hard state. A recent report \citep{b80} claims that IGR J17091$-$3624 is going back to quiescence similar to normal transient BHXBs.  
It is important to note that GRS 1915+105 was discovered as a normal transient in 1992 \citep{b21}. It was detected at the intensity level as low as $\sim$ 90 mCrab and gradually increased up to 300 mCrab over a month period \citep{b20}. Hence it is possible that at the beginning, similar to IGR J17091$-$3624 and normal transients, GRS 1915+105 also evolved from quiescence and passing through the low/hard, HIMS and SIMS, it entered into different variability classes. Following the fate of IGR J17091$-$3624 and other normal transients, GRS 1915+105 may also fade into quiescence in the future. Therefore, it is possible that at the beginning, the spectral evolution of GRS 1915+105 and IGR J17091$-$3624 are similar to normal transients. With the {\it RXTE}/PCA, GRS 1915+105 has usually been seen in the HIMS/SIMS state, but never seen in the typical low hard state. From the HIMS and SIMS state, IGR J17091$-$3624 deviates from the canonical BHXB track and shows regular repetitive variabilities similar to the $\rho$ class variability observed in GRS 1915+105. Besides, within {\it RXTE} era, IGR J17091$-$3624 shows six variability classes \citep{b95} out of 14 classes in GRS 1915+105 \citep{b16,b35}. Thus considering evidences, IGR J17091$-$3624 do show spectral evolution and properties similar to GRS 1915+105, but, during first $\sim$40 days of the outburst, it also clearly exhibits outburst evolution properties similar to other canonical BHXBs (more evidently than GRS 1915+105).    

Our results tentatively indicate that the central object in IGR J17091$-$3624 system may be a black hole. One such indication is the detection of type-B QPOs from six different observations and type-C QPOs from eight different observations (see Table 1). These QPOs are often observed from BHXBs rather than from NSXBs \citep{b81,b86}.
Besides, the observation of low frequency break in the PDS of this source on 15 February, 2011 (also see, \citet{b22}) further decreases the chance of being NSXB. Moreover, from our analysis, type$-$I X-ray bursts are not found in any
observations within about two months. These bursts are unique characteristics of neutron star LMXBs \citep{b53, b52, b34}. This gives an indirect evidence that the source may not be a neutron star. 

We find few dissimilarities between the variable state/$\rho$-like class of IGR J17091$-$3624 and the `$\rho$' class of GRS 1915+105. They are listed below:

\begin{enumerate}

\item The color-color diagram (CD) of the `$\rho$' class shows a loop-like pattern (see \citet{b16}) while the CD of the variable state/$\rho$-like class in IGR J17091$-$3624 shows a patchy pattern (see bottom right panel of Fig. 4). Low count rate and irregular burst structure may cause such randomness in the CD. In the HID, GRS 1915+105 traverses a clock-wise loop while the IGR J17091$-$3624 traverses anti-clockwise \citep{b94}. The X-ray flux from IGR J17091$-$3624 is significantly lower than that observed from GRS 1915+105 even considering a distance as large as $\sim$17 kpc. 

\item An interesting difference between two systems is that the hardness ratio during $\rho$ class activity is higher ($\sim$ 2 times) in IGR J17091$-$3624 than that observed in GRS 1915+105. One possible reason for this higher hardness value, as discussed in \citet{b95}, is the distance of IGR J17091$-$3624 is higher compared to the distance of GRS 1915+105. However, using {\it SWIFT}/XRT data, we find that the absorption column density is 0.9$-$1.02 $\times$ 10$^{22}$ cm$^{-2}$ for IGR J17091$-$3624 while this value is 5$-$13.6 $\times$ 10$^{22}$ cm$^{-2}$ for GRS 1915+105. Hence along with large distance, effects like the presence of the strong disk wind or large disk inclination angle \citep{b96} may also cause the observed flux difference between two sources. The contamination from a nearby source is not favored since the hardening of the lightcurve during the heartbeat oscillations has also been been observed with the {\it SWIFT}/XRT \citep{b98}.

\item Typical low/hard state observations where disk emission is not visible, are commonly seen in canonical BHXBs but have never been seen in GRS 1915+105 during the RXTE era. Earlier outbursts in IGR J17091$-$3624 were started with the LS \citep{b22,b1}. In case of 2011 outburst, {\it SWIFT}/XRT spectral fitting of IGR J17091$-$3624 shows that, from 9 February, 2011 to 20 February, 2011 (see Table 2) disk component is not visible in the spectra. Apart from this, a break in the PDS continuum, detection of broad Lorentzian also indicate that the present outburst started with the LS although accurate determination of spectral state is not possible due to unavailability of {\it RXTE} data.

\item A systematic study of the occurrence of different classes in GRS 1915+105 \citep{b49} show that the $\rho$ class variability eventually arise from the $\alpha$ or $\omega$ class after transiting from the SIMS. These are different X-ray classes with large variation in the variability timescale as well as X-ray flux \citep{b16,b35,b44}. Until now, the $\rho$ class like activity in IGR J17091$-$3624 arises from the SIMS through the IVS, skipping any other variabilities as observed in GRS 1915+105. 

\item From Table 1, the total good time interval of variable state/$\rho$-like class observations in IGR J17091$-$3624 can be estimated as $\sim$19.87 hours. Once in the entire $\rho$ class period (19 March, 2011 to 12 April, 2011), it goes back to the non-variable SIMS (25 March, 2011). During the rise phase of 1997 outburst in GRS 1915+105 also was
in the SIMS prior to $\rho$ class on 26 May, 1997 and
showed a transition from the $\rho$ class to the $\kappa$ class
on 18 June, 1997 \citep{b49}. In GRS 1915+105, the total good time interval of $\rho$ class observations is found to be $\sim$7.36 hours. Unlike IGR J17091$-$3624, GRS 1915+105 occasionally make transition to $\alpha$ or $\kappa$ class within $\rho$ class period \citep{b49}. These observations, in both sources, indicate that `$\rho$' class variability may represent very stable X-ray class. If frequent observations and the total good time interval are assumed to represent the stability of a state, then the possible reason for prolonged stability of IGR J17091$-$3624 in the variable state/$\rho$-like class compared to GRS 1915+105 may be the lower mass accretion rate in IGR J17091$-$3624 compared to GRS 1915+105 which, however, needed to be explored further.

\end{enumerate}

Using the radio data, \citet{b92} found the distance of IGR J17091$-$3624 to be 10$-$17 kpc assuming black hole mass to be 10 M$_{\odot}$. Later, \citet{b85} estimated the mass of 6 M$_{\odot}$. We assume here the black hole mass to be 8 $\pm$ 2 M$_{\odot}$. We use the distance of IGR J17091$-$3624 to be 14$\pm$3 kpc to calculate the mass accretion rate of IGR J17091$-$3624 relative to the GRS 1915+105. In case of GRS 1915+105, we assume the mass, the distance and the disk inclination angle with respect to observer's line of sight to be 10.1$\pm$0.4 M$_{\odot}$ \citep{b79}, 12.5$\pm$2.1 kpc and 70$^o$ \citep{b89} respectively. The mass accretion rate at the inner disk can be derived using the energy conservation law and the Virial theorem. Using simple analytical approach and assuming the disk radiation to be blackbody \& considering both side of the disk, the mass accretion rate is \citep{b23} :

\begin{equation}
\dot{m} = 8 \pi R_{in,km}^3 \sigma T_{in,keV}^4 / 3GM_{bh} ~~ M_{\odot} yr^{-1}
\end{equation}

where $\sigma$ is the Boltzmann constant, G is the gravitational constant and T$_{in,keV}$ is the inner disk temperature in keV and R$_{in,km}$ is given by the equation

\begin{equation}
R_{in,km} = C^2 \times \sqrt{N_{\it diskbb}} \times D_{10,kpc}/\sqrt{\cos i}
\end{equation}

where C is the color correction factor (assumed to be $\sim$ 1.8 in case of black hole \citep{b84}), N$_{\it diskbb}$ is the normalization corresponds to the {\it diskbb} model component in {\tt XSpec}, D$_{10,kpc}$ is the distance to the source in the unit of 10 kpc and {\it i} is the disk inclination angle with respect to observer's line of sight. Inserting the equation (2) in (1), we get

\begin{equation}
\dot{m} = \frac{ 8 \pi \sigma T_{in,keV}^4 }{ 3GM_{bh}} \times (C^2 \times \sqrt{ N_{\it diskbb}} \times D_{10,kpc}/\sqrt{\cos i})^3
\end{equation}

To obtain a qualitative idea, we compare accretion parameters of both sources using above equations. For example, the spectral analysis of a less-variable, HIMS observation in GRS 1915+105 on 09 September, 1997 using {\it RXTE}/PCA yields the disk temperature of 1.24 $\pm$ 0.03 keV and inner disk radius of 57 $\pm$ 3 km \citep{b89}. Using equation (2), the equivalent normalization would be  211.3 $\pm$ 26.5. From the spectral analysis of non-variable, HIMS in IGR J17091$-$3624, observed on 08 March, 2011, we find the value of N$_{\it diskbb}$ to be 13.87 $\pm$ 1.35 and the disk temperature to be 1.24 $\pm$ 0.03 keV (See Table 2). Using these values in equation (3) and assuming disk inclination angle (i) to be 70$^o$ in IGR J17091$-$3624 \citep{b96,b98}, we find $\dot{\bf m}$$_{GRS 1915+105}$/$\dot{\bf m}$$_{IGR J17091-3624}$ $\sim$ 48.3 $\pm$ 8.7. Hence, the mass accretion rate in GRS 1915+105 is significantly higher (may be in the order of magnitude) compared to that of IGR J17091$-$3624. This may be consistent with very high X-ray flux observed in GRS 1915+105 than that seen from IGR J17091$-$3624 even considering IGR J17091$-$3624 at larger distance. We also calculate the relative viscous timescale of the accretion flow using the following relation \citep{b23}:
\begin{equation}
t_{vis}^d = 4.3 \times 10^{-4} \alpha^{-1} \dot{m}^{-1}_d M_{bh}^{-1} R_{in,km}^2 ~~ s
\end{equation}
where t$_{vis}^d$ is the dynamic viscous timescale, $\alpha$ is the viscosity parameter. We consider the typical value of $\alpha$ to be 0.01 for both sources. Using relation (4) and above parameters, we find the value of t$_{vis|GRS 1915+105}^d$/t$_{vis|IGR J17091-3624}^d$ to be $\approx$ 0.37 $\pm$ 0.13 when the disk inclination angle of IGR J17091$-$3624 is $\sim$70$^o$ \citep{b96,b98}. Hence the viscous timescale of GRS 1915+105 is smaller relative to IGR J17091$-$3624 in non-variable state.

\section{Conclusions}\label{Conclusion}

From our analysis of {\it SWIFT}/XRT and {\it RXTE}/PCA data during 2011 outburst of IGR J17091$-$3624, 
and using results from comparative study of timing and spectral properties with GRS 1915+105, 
we conclude following points.

\begin{enumerate}

\item We find that IGR J17091$-$3624 is the only known transient LMXB, which shows regular, repetitive and structured variability in the intensity pattern similar to the $\rho$-type variability seen from GRS 1915+105. Various parameters like burst frequency evolution, burst structure profile, rise and decay profile, peak-to-minimum flux ratio, harmonics of mHz characteristic pulsations and PDS characteristics show remarkable similarity. The entry to the $\rho$-type variability in IGR J17091$-$3624 is from the SIMS via intermediate variability state which show $\sim$ 11-14 mHz coherent pulsations with the presence of weak red noise/type-A QPO, while the exit from the $\rho$-type variability is through the soft state with several broad Lorentzian noise components and QPO-like feature at $\sim$ 3 Hz (similar to the $\kappa$/$\lambda$ class; see Table 1). GRS 1915+105 also was in the SIMS prior to $\rho$ class on 26 May, 1997 and
showed a transition from the $\rho$ class to the $\kappa$ class
on 18 June, 1997 \citep{b49}. However, significant differences in hardness ratio as well as variance spectra are also observed.  

\item Several evidences like power-law dominated spectra, the detection of type-B and type-C QPOs, break in PDS continuum along with $\rho$ class activity (as seen in GRS 1915+105 which is a BHXB) are consistent with the black hole identification of the source.

\item The source shows increase in flux while transiting from the low/hard state/HIMS to the SIMS. Later a transition occurs from a typical irregular variability state (SIMS), seen commonly in most of the BHXBs to the regular, repetitive large variability state, seen only in GRS 1915+105 (`$\rho$'-type variability) via an intermediate state. However, X-ray bursts during variability state/$\rho$-like class are spectrally harder than that found in GRS 1915+105. Central black hole mass, its distance, disk inclination angle, absorption column density etc. may be important to explain the observed hardness ratio but at present we do not know exact answer. It is a topic for future investigations if these parameters or some unknown parameter/mechanism is responsible for higher hardness ratio during $\rho$ class activity in IGR J17091$-$3624.

\item From Table 1, it is clear that we detect several QPOs ranging from few mHz to 10 Hz or so. Not only that, there are to and fro transition between the HIMS and the SIMS and the source show evolution of characteristic oscillations in $\rho$-class faster than GRS 1915+105.  
 
\item In both sources, `$\rho$' class activity is observed following the soft intermediate state where disk emission in the energy spectra is visible and contribute significantly to the total flux. GRS 1915+105 during the {\it RXTE} era, never shows typical low/hard state where disk emission is not visible but IGR J17091$-$3624 shows low hard state where disk emission is not significant. This result agrees well with the low hard state observations of other black hole X-ray binaries. Similar to a normal transient, IGR J17091$-$3624 also seems to move to the quiescence as observed with {\it SWIFT}/XRT \citep{b80}. Hence considering evidences, IGR J17091$-$3624, behaves similar to GRS 1915+105 at certain period of evolution whereas it behaves like other canonical BHXBs at some other period of evolution. Hence, this source may be an important bridge between GRS 1915+105 and other canonical BHXBs in understanding observational dissimilarities, as well as accretion and radiation mechanisms.

\end{enumerate}

Thus long term and continuous observation of outbursts of this source in multi-wavelength band using {\it ASTROSAT} would reveal a lot of information regarding the radiation mechanism and accretion flow properties of BHXBs, and help to construct the complete picture of accretion flow around black hole systems by connecting GRS 1915+105 with other canonical BHXB systems.

\section{Acknowledgements}

We thank the referee for his/her constructive comments which helps to improve the work. We also thank another referee Ron Remillard for usuful discussions. MP is thankful to Ranjeev Misra for constructive discussions. We thank T. Belloni for providing {\tt GHATS v1.0.2} timing analysis software. This research has made use of the General High-energy Aperiodic Timing Software (GHATS) package developed by T.M. Belloni at INAF - Osservatorio Astronomico di Brera and data obtained through the High Energy Astrophysics Science Archive Research Center online service, provided by the NASA/Goddard Space Flight Center and the {\it SWIFT} data center.

\clearpage

\begin{table*}
\centering
 \caption{\scriptsize{Observational details of IGR J17091$-$3624 using {\it RXTE}/PCA and {\it SWIFT}/XRT data during 2011 outburst. }} 
\begin{center}
\scalebox{0.70}{%
\begin{tabular}{|l|l|l|l|l|l|l|l|l|l|}
 \hline
\hline
 Observation & MJD & Instrument & Observation & Good time & Average      & Average  & spectral   & PDS features \\
 date  (DD-MM-YY) &    &            & ID          &  interval (sec)   &  soft color  & hard color   &  state/class &   \\ 
\hline
\hline
09-02-11 & 55601.06 & XRT(WT) & 00031921005 & 1467.4 & -- & -- &  LS? & band-limited noise observed  \\
\hline
10-02-11 & 55602.58 & XRT(WT) & 00031921006 & 2200.4 & -- & -- & LS? & strong band-limited noise (rms amplitude (\%) 9.4 $\pm$ 0.6)  \\
\hline
12-02-11 & 55604.21 & XRT(WT) & 00031921008 & 2198.1 & -- & -- & LS? & band-limited noise observed \\
\hline
13-02-11 & 55605.21 & XRT(WT) & 00031921009 & 2151.2 & -- & -- &  LS? & band-limited noise observed \\
\hline
14-02-11 & 55606.16 & XRT(WT) & 00031921010 & 1877.6 & -- & -- & LS? & band limited noise observed \\
\hline
15-02-11  & 55607.23 & XRT(WT) & 00031921011 & 2101.6 & -- & -- & LS &  band-limited noise \& a break in PDS is observed ($\sim$ 0.19$\pm$0.03 Hz)\\
\hline
16-02-11 & 55608.23 & XRT(WT)  & 00031921012  &  2182.3 & -- &  --  & LS & band-limited noise \& broad QPO-like\\
& & & & & & & & and feature observed  (0.18$\pm$0.02 Hz, 2.37$\pm$0.18, 13.9$\pm$2.6)\\
\hline
18-02-11 & 55610.17 & XRT(WT) & 00031921013  & 2176.5 & -- & -- & LS & strong band-limited noise \& broad QPO-like\\
& & & & & & & & and feature observed (0.22$\pm$0.03 Hz, 2.25$\pm$0.18, 8.7$\pm$1.3)\\
\hline
20-02-11 & 55612.25 & XRT(WT) & 00031921014 & 2104.2 & -- & -- & HIMS & pds fitted with ({\tt lore+bknpower})(122.1/125), \\
& & & & & & & & type-C QPO (0.51$\pm$0.13 Hz, 5.25$\pm$0.12, 12.8$\pm$0.6) \\ 
& & & & & & & & and flat top noise ($\sim$ 7\%) detected \\
\hline
22-02-11 & 55614.19 & XRT(WT) & 00031921015 & 2050.2 & -- & -- & SIMS & broad Lorentzian ($\sim$1.89 Hz) and weak red noise detected \\
\hline
23-02-11 & 55615.99 & PCA & 96065-03-01-03 & 3184.4 & 0.73$\pm$0.01 & 0.36$\pm$0.02 & SIMS & pds fitted with ({\tt lore+bknpower}(225.9/224)), \\
& & & & & & & & transient type-B QPO (4.16$\pm$0.08 mHz, 5.71$\pm$0.34, 6.68$\pm$0.79)\\	
& & & & & & & & with weak red noise (rms amplitude(\%) 4.25$\pm$0.18)\\
\hline
24-02-11 & 55616.25 & XRT(WT) & 00031921016 & 1065.6 & -- & -- & SIMS & pds fitted with ({\tt lore+lore+bknpower})(117.3/122),\\
& & & & & & & & transient type-B QPO (4.32$\pm$0.37 Hz, 7.17$\pm$0.36, 5.28$\pm$0.32)\\	
& & & & & & & & broad Lorentzian noise (0.83$\pm$0.14 Hz, 7.82$\pm$0.27)\\

\hline
26-02-11 & 55619.01 & PCA & 96065-03-02-00 & 1712.2 & 0.75$\pm$0.01 & 0.41$\pm$0.02 & HIMS & pds fitted with ({\tt lore+bknpower}(213.5/224)),\\
& & & & & & & & type-C QPO (3.73$\pm$0.09 Hz, 4.76$\pm$0.14, 16.9$\pm$0.6)\\	
& & & & & & & & with strong flat-top noise (rms amplitude(\%) 7.85$\pm$0.18)\\

\hline 
28-02-11 & 55621.34 & XRT(WT) & 00031921017 & 2576.6 & -- & -- & SIMS & pds fitted with ({\tt lore+bknpower})(121.8/125),\\
& & & & & & & & transient type-B QPO (3.84$\pm$0.11 Hz, 6.79$\pm$0.23, 5.31$\pm$0.55)\\	
& & & & & & & & with weak red noise (rms amplitude(\%) 3.96$\pm$0.49)\\
\hline
02-03-11 & 55622.59 & PCA & 96420-01-01-00 & 3744.3 & 0.72$\pm$0.04 & 0.39$\pm$0.08 & HIMS & pds fitted with ({\tt lore+bknpower}(234.2/224)),\\
& 55622.52 & XRT(WT) & 00031921018 & 2340.9 & -- & -- & & type-C QPO (3.67$\pm$0.26 Hz, 4.17$\pm$0.56, 21.3$\pm$0.5) \\
& & & & & & & & with strong flat-top noise (rms amplitude(\%) 8.19$\pm$0.47)\\

\hline
03-03-11 & 55623.28 & PCA & 96420-01-01-01 & 2064.5 & 0.68$\pm$0.03 & 0.36$\pm$0.02 & SIMS & pds fitted with ({\tt lore+bknpower}(229.9/224)), \\
& 55623.86 & PCA & 96420-01-01-02 & 6944.4 & 0.66$\pm$0.02 & 0.38$\pm$0.03 & & transient type-B QPO (4.85$\pm$0.08 Hz, 7.67$\pm$0.68, 6.14$\pm$0.29) \\
& 55623.29 & XRT(WT) & 00031921019 & 971.4 & -- & -- & & with weak red noise (rms amplitude(\%) 3.43$\pm$0.59)\\

\hline
04-03-11 & 55624.56 & PCA & 96420-01-02-00 & 1584.2 & 0.62$\pm$0.03 & 0.33$\pm$0.01 & SIMS to & pds fitted with ({\tt lore+bknpower}(212.6/224)),\\
& 55624.85 & XRT(WT) & 00031921020 & 2289.6 & -- & -- & HIMS & type-B/type-C QPO (5.32$\pm$0.46 Hz, 6.38$\pm$0.56, 6.31$\pm$0.37) \\
& & & & & & & transition & with flat-top noise (rms amplitude(\%) 6.58$\pm$0.79)\\

\hline
06-03-11 & 55626.39 & PCA & 96420-01-02-01 & 65.5 & 0.67$\pm$0.03 & 0.35$\pm$0.01 & HIMS & pds fitted with ({\tt lore+bknpower}(217.3/224)),\\
& 55626.23 & XRT(WT) & 00031921022 & 1704.1 & -- & -- & & transient type-C QPO (5.15$\pm$0.11 Hz, 4.98$\pm$0.27, 9.11$\pm$0.09) \\
& & & & & & & & with strong flat-top noise (rms amplitude(\%) 7.12$\pm$0.35)\\

\hline
08-03-11 & 55628.47 & PCA & 96420-01-02-02 & 3202.2 & 0.64$\pm$0.01 & 0.36$\pm$0.01 & HIMS & pds fitted with ({\tt lore+bknpower}(226.4/224)),\\
& 55628.23 & XRT(WT) & 00031921022 & 1704.1 & -- & -- & & transient type-C QPO (5.40$\pm$0.23 Hz, 4.85$\pm$0.39, 8.87$\pm$0.69) \\
& & & & & & & & with strong flat-top noise (rms amplitude(\%) 7.12$\pm$0.35)\\
       
\hline
10-03-11 & 55630.84 & PCA & 96420-01-02-03 & 1648.8 & 0.67$\pm$0.02 & 0.37$\pm$0.04 & HIMS & pds fitted with ({\tt lore+bknpower}(218.1/224)),\\
& 55630.16 & XRT(WT) & 00031921023 & 1401.5 & -- & -- & & transient type-C QPO (5.23$\pm$0.09 Hz, 5.31$\pm$0.23, 10.9$\pm$0.7) \\
& & & & & & & & with strong flat-top noise (rms amplitude(\%) 8.89$\pm$0.49)\\

\hline       
12-03-11 & 55632.54 & PCA & 96420-01-03-00 & 1520.4 & 0.63$\pm$0.02 & 0.33$\pm$0.02 & SIMS & pds fitted with ({\tt lore+bknpower}(229.6/224)), \\
& 55632.73 & XRT(WT) & 00031921024 & 2182.3 & -- & -- & & narrow, weak QPO-like feature (0.68$\pm$0.03 Hz, 12.5$\pm$0.4, 3.08$\pm$0.11) \\
& & & & & & & & with weak red noise (rms amplitude(\%) 5.15$\pm$0.26)\\
         
\hline
13-03-11 & 55633.24 & XRT(WT) & 00031921025 & 2009.8 & -- & -- & SIMS & no QPO is detected and \\
         & & & & & & &    &                                                       weak red noise observed (rms amplitude(\%) 2.55$\pm$0.34)\\
\hline
14-03-11 & 55634.72 & PCA & 96420-01-03-01 & 3344.2 & 0.53$\pm$0.02 & 0.28$\pm$0.06 & IVS & pds fitted with ({\tt lore+lore+bknpower})(231.8/221),\\
& & & & & & & & characteristic QPO (11.5$\pm$0.8 mHz, 4.46$\pm$0.58, 5.72$\pm$0.18),\\	
& & & & & & & & narrow, weak QPO-like feature (2.61$\pm$0.12 Hz, 11.8$\pm$0.7, 2.97$\pm$0.23)\\
& & & & & & & & with weak red noise (rms amplitude(\%) 3.69$\pm$0.25)\\
	 								
\hline
15-03-11 & 55635.65 & XRT(WT) & 00031921026 & 1481.2 & -- & -- & SIMS & pds fitted with ({\tt lore+bknpower})(118.3/125),\\
& & & & & & & & transient type-B QPO (4.46$\pm$0.21 Hz, 6.09$\pm$0.53, 7.81$\pm$0.86)\\	
& & & & & & & & with weak red noise (rms amplitude(\%) 2.99$\pm$0.35)\\

\hline
18-03-11 & 55638.54 & XRT(WT) & 00031921029 & 2164.6 & -- & -- & IVS & pds fitted with ({\tt lore+bknpower})(114.6/125),\\
& & & & & & & & characteristic QPO (13.3$\pm$1.4 mHz, 8.6$\pm$2.3, 7.39$\pm$0.41)\\	
& & & & & & & & with weak red noise (rms amplitude(\%) 3.65$\pm$0.33)\\                         
\hline
19-03-11 & 55639.71 & PCA & 96420-01-04-00 & 2752.3 & 0.50$\pm$0.01 & 0.21$\pm$0.01 & variable & pds fitted with ({\tt lore+lore+bknpower})(212.6/221),\\
& 55639.67 & XRT(WT) & 00031921028 & 2631.5 & -- & -- & state/ & heartbeat $\nu_f$ (29.6$\pm$1.2 mHz, 7.42$\pm$0.51, 28.6$\pm$1.8)\\
& & & & & & & $\rho$ class & heartbeat $\nu_{h1}$ (62.4$\pm$4.5 mHz, 5.56$\pm$0.36, 9.02$\pm$0.48)\\
& & & & & & & & heartbeat $\nu_{h2}$ (0.11$\pm$0.05 Hz, 5.77$\pm$0.42, 4.26$\pm$0.53)\\
         
\hline
20-03-11 & 55640.47 & XRT(WT) & 00031921030 & 2352.5 & -- & -- & variable & pds fitted with ({\tt lore+lore+lore+bknpower})(223.6/218),\\
& & & & & & & state/ & heartbeat $\nu_f$ (31.2$\pm$2.3 mHz, 18.6$\pm$2.2, 25.9$\pm$1.8)\\	
& & & & & & & $\rho$ class & heartbeat $\nu_{h1}$ (63.4$\pm$4.8 mHz, 4.84$\pm$0.24, 7.77$\pm$0.83)\\
& & & & & & & & heartbeat $\nu_{h2}$ (0.11$\pm$0.08 Hz, 2.77$\pm$0.12, 6.45$\pm$0.34)\\
\hline
\hline
\end{tabular}}
\end{center}
\end{table*}
\clearpage

\begin{table*}
\centering
\tablecomments {{\scriptsize Continuation of Table 1 ...}}
\begin{center}
\scalebox{0.70}{%
\begin{tabular}{|l|l|l|l|l|l|l|l|l|}
 \hline
\hline
 Observation & MJD & Instrument & Observation & Good time         & Average      & Average      & spectral  & PDS features\\
  date  (DD-MM-YY) &    &            & ID          &  interval (sec)   &  soft color  & hard color & state/class &            \\ 
\hline

\hline
22-03-11 & 55642.20 & PCA & 96420-01-04-02 & 2976.4 & 0.54$\pm$0.01 & 0.29$\pm$0.09 & variable & pds fitted with ({\tt lore+lore+lore+bknpower})(228.6/218),\\
         & 55642.35 & XRT(WT) & 00031921031 & 2260.3 & -- & -- & state/ & heartbeat $\nu_f$ (25.5$\pm$1.4 mHz, 9.69$\pm$0.74, 22.0$\pm$1.6)\\
& & & & & & & $\rho$ class & heartbeat $\nu_{h1}$ (51.7$\pm$2.4 mHz, 5.29$\pm$0.33, 9.39$\pm$0.88)\\
& & & & & & & & transient type-B QPO (2.73$\pm$0.15 Hz, 8.36$\pm$2.43, 3.16$\pm$0.22)\\

\hline
23-03-11 & 55643.78 & PCA & 96420-01-04-01 & 1152.4 & 0.55$\pm$0.11 & 0.34$\pm$0.03 & IVS & pds fitted with ({\tt lore+bknpower})(233.4/224),\\
& & & & & & & & characteristic QPO (12.7$\pm$1.1 mHz, 11.6$\pm$1.6, 6.87$\pm$0.49)\\	
& & & & & & & & with weak red noise (rms amplitude(\%) 2.67$\pm$0.19)\\

\hline
24-03-11 & 55644.75 & PCA & 96420-01-04-03 & 2752.8 & 0.57$\pm$0.03 & 0.33$\pm$0.04 & IVS & pds fitted with ({\tt lore+lore+lore+bknpower})(227.5/218),\\
& & & & & & & & $\nu_f$ (11.6$\pm$0.9 mHz, 2.89$\pm$0.16, 6.02$\pm$0.39)\\	
& & & & & & &  & broad Lorentzian noise (60.1$\pm$3.7 mHz, 4.15$\pm$0.21) and \\
& & & & & & & & transient type-A QPO (8.18$\pm$0.62 Hz, 3.75$\pm$0.29, 5.26$\pm$0.18)\\

\hline
25-03-11 & 55645.86 & PCA & 96420-01-05-02 & 3312.3 & 0.58$\pm$0.01 & 0.37$\pm$0.05 & SIMS & pds fitted with ({\tt lore+bknpower})(214.4/224),\\
& & & & & & & & transient type-A QPO (7.78$\pm$0.45 Hz, 4.28$\pm$0.87, 4.57$\pm$0.29)\\	
& & & & & & & & with weak red noise (rms amplitude(\%) 4.13$\pm$1.16)\\

\hline
26-03-11 & 55646.89 & XRT(WT) & 00031921034 & 2160.1 & -- & -- & IVS & pds fitted with ({\tt lore+bknpower})(123.8/125),\\
& & & & & & & & characteristic QPO (14.1$\pm$0.2 mHz, 8.97$\pm$1.1, 6.10$\pm$0.66)\\	
& & & & & & & & with weak red noise (rms amplitude(\%) 3.23$\pm$0.26)\\

\hline
27-03-11 & 55648.01 & PCA & 96420-01-05-00 & 32990.4 & 0.52$\pm$0.04 & 0.27$\pm$0.03 & variable & pds fitted with ({\tt lore+lore+lore+bknpower})(228.4/218),\\
& & & & & & & state/ & heartbeat $\nu_f$ (21.1$\pm$1.4 mHz, 14.8$\pm$1.1, 13.3$\pm$1.2)\\	
& & & & & & & $\rho$ class & heartbeat $\nu_{h1}$ (44.2$\pm$2.9 mHz, 13.3$\pm$1.2, 4.89$\pm$0.57)\\
& & & & & & & & heartbeat $\nu_{h2}$ (85.6$\pm$5.5 mHz, 16.8$\pm$1.5, 6.15$\pm$0.74)\\

\hline
29-03-11 & 55649.06 & PCA & 96420-01-05-03 & 2640.3 & 0.55$\pm$0.01 & 0.33$\pm$0.02 & variable & pds fitted with ({\tt lore+lore+bknpower})(232.7/221),\\
& & & & & & & state/ & heartbeat $\nu_f$ (23.8$\pm$1.9 mHz, 5.39$\pm$0.69, 15.9$\pm$1.4)\\	
& & & & & & & $\rho$ class & heartbeat $\nu_{h1}$ (49.7$\pm$2.2 mHz, 11.8$\pm$3.6, 6.89$\pm$0.56)\\

\hline
30-03-11 & 55650.98 & PCA & 96420-01-05-01 & 2304.5 & 0.48$\pm$0.02 & 0.28$\pm$0.02 & variable & pds fitted with ({\tt lore+lore+lore+bknpower})(213.5/218),\\
         & 55650.74 & XRT(WT) & 00031921036 & 331.65 & -- & -- & state/ & heartbeat $\nu_f$ (24.1$\pm$1.6 mHz, 13.5$\pm$0.3, 15.9$\pm$1.4)\\ 
& & & & & & & $\rho$ class & heartbeat $\nu_{h1}$ (50.6$\pm$2.8 mHz, 16.8$\pm$0.6, 7.88$\pm$0.71)\\
& & & & & & & & heartbeat $\nu_{h2}$ (75.7$\pm$5.3 mHz, 9.54$\pm$0.93, 5.06$\pm$0.48)\\

\hline
31-03-11 & 55651.88 & PCA & 96420-01-05-04 & 3968.1 & 0.51$\pm$0.02 & 0.35$\pm$0.04 & variable & pds fitted with ({\tt lore+lore+lore+bknpower})(222.8/218),\\
& & & & & & & state/ & heartbeat $\nu_f$ (26.3$\pm$1.4 mHz, 14.1$\pm$0.5, 16.8$\pm$1.7)\\	
& & & & & & & $\rho$ class & heartbeat $\nu_{h1}$ (52.8$\pm$3.9 mHz, 12.7$\pm$0.3, 7.53$\pm$1.05)\\
& & & & & & & & heartbeat $\nu_{h2}$ (79.6$\pm$4.4 mHz, 9.41$\pm$0.23, 9.19$\pm$1.16)\\
& & & & & & & & transient type-A QPO (8.36$\pm$0.26 Hz, 4.79$\pm$0.38, 3.66$\pm$0.31)\\

\hline
02-04-11 & 55653.70 & PCA & 96420-01-06-00 & 3488.7 & 0.50$\pm$0.01 & 0.29$\pm$0.01 & variable & pds fitted with ({\tt lore+lore+lore+bknpower})(226.5/218),\\
& & & & & & & state/ & heartbeat $\nu_f$ (33.6$\pm$1.3 mHz, 11.2$\pm$0.3, 22.3$\pm$1.3)\\	
& & & & & & & $\rho$ class & heartbeat $\nu_{h1}$ (67.8$\pm$2.6 mHz, 5.84$\pm$0.21, 10.9$\pm$2.6)\\
& & & & & & & & heartbeat $\nu_{h2}$ (0.11$\pm$0.01 Hz, 3.19$\pm$0.17, 6.23$\pm$0.13)\\

\hline
03-04-11 & 55654.89 & PCA & 96420-01-06-01 & 6112.3 & 0.53$\pm$0.01 & 0.31$\pm$0.04 & variable & pds fitted with ({\tt lore+lore+lore+lore+bknpower})(207.4/215),\\
& & & & & & & state/ & heartbeat $\nu_f$ (36.6$\pm$0.7 mHz, 9.95$\pm$0.54, 29.3$\pm$1.3)\\	
& & & & & & & $\rho$ class & heartbeat $\nu_{h1}$ (75.9$\pm$4.1 mHz, 4.21$\pm$0.19, 15.3$\pm$1.1)\\
& & & & & & & & broad Lorentzian noise (0.21$\pm$0.05 Hz, 9.83$\pm$0.79) \\
& & & & & & & & transient type-A QPO (7.28$\pm$0.14 Hz, 3.21$\pm$0.17, 4.89$\pm$0.14)\\

\hline
05-04-11 & 55656.71 & PCA & 96420-01-06-02 & 5136.3 & 0.54$\pm$0.01 & 0.27$\pm$0.04 & variable & pds fitted with ({\tt lore+lore+bknpower})(231.5/221),\\
& & & & & & & state/ & heartbeat $\nu_f$ (39.1$\pm$1.6 mHz, 6.89$\pm$0.32, 32.5$\pm$2.2)\\	
& & & & & & & $\rho$ class & heartbeat $\nu_{h1}$ (81.3$\pm$3.8 mHz, 3.95$\pm$0.23, 14.9$\pm$1.4)\\

\hline
06-04-11 & 55657.31 & PCA & 96420-01-06-03 & 1812.8 & 0.55$\pm$0.01 & 0.25$\pm$0.02 & variable & pds fitted with ({\tt lore+lore+lore+bknpower})(225.8/218),\\
& & & & & & & state/ & heartbeat $\nu_f$ (35.8$\pm$2.1 mHz, 4.56$\pm$0.19, 14.5$\pm$1.6)\\	
& & & & & & & $\rho$ class & heartbeat $\nu_{h1}$ (64.8$\pm$4.7 mHz, 4.80$\pm$0.13, 23.6$\pm$1.1)\\
& & & & & & & & transient type-B QPO (4.02$\pm$0.12 Hz, 2.15$\pm$0.17, 13.9$\pm$0.8)\\

\hline
10-04-11 & 55661.75 & PCA & 96420-01-07-00 & 3376.6 & 0.58$\pm$0.01 & 0.28$\pm$0.03 & variable & pds fitted with ({\tt lore+lore+lore+bknpower})(230.3/218),\\
& & & & & & & state/ & heartbeat $\nu_f$ (45.9$\pm$2.1 mHz, 8.47$\pm$0.76, 33.8$\pm$1.6)\\	
& & & & & & & $\rho$ class & heartbeat $\nu_{h1}$ (94.2$\pm$3.4 mHz, 9.33$\pm$0.36, 11.6$\pm$1.7)\\
& & & & & & & & transient type-A QPO (7.76$\pm$0.35 Hz, 2.48$\pm$0.22, 7.53$\pm$0.45)\\

\hline
11-04-11 & 55662.56 & PCA & 96420-01-07-01 & 3375.2 & 0.56$\pm$0.01 & 0.27$\pm$0.02 & variable & pds fitted with ({\tt lore+lore+lore+bknpower})(210.7/218),\\
& & & & & & & state/ & heartbeat $\nu_f$ (45.2$\pm$0.7 mHz, 4.61$\pm$0.25, 31.8$\pm$2.2)\\	
& & & & & & & $\rho$ class & heartbeat $\nu_{h1}$ (90.4$\pm$3.5 mHz, 3.22$\pm$0.25, 11.2$\pm$0.9)\\
& & & & & & & & broad Lorentzian noise (5.52$\pm$1.38 Hz, 9.98$\pm$1.13) \\

\hline
12-04-11 & 55663.34 & PCA & 96420-01-07-02 & 3278.5 & 0.58$\pm$0.01 & 0.28$\pm$0.02 & variable & pds fitted with ({\tt lore+lore+lore+bknpower})(224.4/218),\\
& & & & & & & state/ & heartbeat $\nu_f$ (44.5$\pm$0.6 mHz, 2.69$\pm$0.13, 30.4$\pm$1.7)\\
& & & & & & & $\rho$ class & heartbeat $\nu_{h1}$ (90.1$\pm$0.2 mHz, 13.8$\pm$1.6, 8.63$\pm$0.73)\\
& & & & & & & & transient type-B QPO (4.93$\pm$0.09 Hz, 2.46$\pm$0.22, 4.39$\pm$0.11)\\
	
\hline
15-04-11 & 55666.55 & PCA & 96420-01-08-00 & 1513.6 & 0.59$\pm$0.03 & 0.29$\pm$0.03 & $\rho$ to other & pds fitted with ({\tt lore+lore+lore+bknpower})(211.6/218), \\
& & & & & & &  class transition & broad Lorentzian noise (56.73$\pm$3.43 mHz, 30.1$\pm$2.4), \\
& & & & & & & (may be $\kappa$/$\lambda$) & broad Lorentzian noise (0.22$\pm$0.07 Hz, 19.4$\pm$1.5) and \\
& & & & & & & & broad QPO-like feature (3.09$\pm$0.05 Hz, 3.25$\pm$0.36, 9.65$\pm$1.23) \\ 
	 
\hline 
\end{tabular}}
\hspace{2em}
\tablecomments{LS, HIMS, SIMS \& IVS stand for the low/hard state, Hard intermediate state, soft intermediate state and Intermediate variable state respectively. In the last column, quantities in braces for quasi-periodic oscillations (QPOs) represent (frequency with error, q-factor with error, rms amplitude with error) and for noise components (frequency with error, rms with error). $\nu_f$, $\nu_{h1}$ and $\nu_{h2}$ represent fundamental QPO of $\rho$-like/variable state oscillations, its first and second harmonics respectively. Models used to fit PDS and $\chi^2$/dof are also provided.}
\end{center}
\end{table*}
\clearpage

\begin{table*}
\centering
 \caption{Best fit parameters (with 1$\sigma$ error bars) obtained by fitting the low/hard state, hard intermediate and soft intermediate state spectra of IGR J17019$-$3624 with the {\tt powerlaw} and the {\tt diskbb+powerlaw} models respectively.} 
\begin{center}
\scalebox{0.62}{%
\begin{tabular}{|l|l|l|l|l|l|l|l|l|l|l|l|l|l|}
 \hline
\hline
Instrument  & Observation  & Date & state/  & $\Gamma$   &  kT$_{in}$ & N$_{diskbb}$ & Flux$_{total}$ & Flux$_{powerlaw}$ & Flux$_{diskbb}$ & $\chi^2$/dof \\
 used &            ID            & (DD-MM-YY)& class     & & (keV) & & & & & \\
\hline
 XRT & 00031921005 & 09-02-11 & LS? & 1.66$^{+0.05}_{-0.04}$ & -- & -- & 0.61 & -- & -- & 657.98/632  \\
\hline
 XRT & 00031921006 & 10-02-11 & LS? & 1.75$^{+0.05}_{-0.04}$ & -- & -- & 0.45 & -- & -- & 573.58/632  \\
\hline
 XRT & 00031921008 & 12-02-11 & LS? & 1.63$^{+0.03}_{-0.03}$ & -- & -- & 0.87 & -- & -- & 728.57/632  \\
\hline
 XRT & 00031921009 & 13-02-11 & LS? & 1.69$^{+0.03}_{-0.04}$ & -- & -- & 0.98 & -- & -- & 667.64/632  \\
\hline
 XRT & 00031921010 & 14-02-11 & LS? & 1.73$^{+0.04}_{-0.05}$ & -- & -- & 0.71 & -- & -- & 694.01/632  \\
\hline
 XRT & 00031921011 & 15-02-11 & LS & 1.67$^{+0.03}_{-0.04}$ & -- & -- & 0.95 & -- & -- & 726.74/632  \\
\hline
 XRT & 00031921012 & 16-02-11 & LS & 1.62$^{+0.03}_{-0.03}$ & -- & -- & 1.11 & -- & -- & 629.34/632  \\
\hline
 XRT & 00031921013 & 18-02-11 & LS & 1.66$^{+0.03}_{-0.04}$ & -- & -- & 1.03 & -- & -- & 586.28/632  \\
\hline
 XRT & 00031921014 & 20-02-11 & HIMS & 1.79$^{+0.04}_{-0.05}$ & -- & -- & 0.98 & -- & -- & 609.43/632  \\
\hline
 XRT & 00031921015 & 22-02-11 & SIMS & 2.22$^{+0.07}_{-0.05}$ & 0.96$^{+0.04}_{-0.06}$ & 21.7$^{+8.8}_{-4.6}$ & 2.47 & 1.85 & 0.63 & 595.40/632  \\
\hline
PCA & 96065-03-01-03 & 23-02-11 & SIMS & 2.29$^{+0.03}_{-0.03}$ & 1.10$^{+0.04}_{-0.06}$ & 6.43$^{+2.42}_{-1.45}$ & 0.79 & 0.68 & 0.11 & 45.12/42  \\
\hline
PCA & 96065-03-02-00 & 26-02-11 & SIMS & 2.38$^{+0.04}_{-0.07}$ & 0.93$^{+0.11}_{-0.06}$ & 11.50$^{+4.30}_{-5.50}$ & 0.68 & 0.59 & 0.08 & 43.59/42  \\ 
\hline
PCA & 96420-01-01-00 & 02-03-11 & HIMS & 2.22$^{+0.02}_{-0.03}$ & 1.10$^{+0.06}_{-0.04}$ & 2.69$^{+0.76}_{-0.52}$ & 0.47 & 0.43 & 0.04 & 44.52/42  \\
\hline
PCA & 96420-01-01-01 & 03-03-11 & SIMS & 2.30$^{+0.08}_{-0.03}$ & 1.14$^{+0.04}_{-0.03}$ & 4.36$^{+0.56}_{-0.48}$ & 0.49 & 0.40 & 0.09 & 39.42/42  \\
\hline
PCA & 96420-01-01-02 & 03-03-11 & SIMS & 2.41$^{+0.04}_{-0.03}$ & 0.82$^{+0.10}_{-0.09}$ & 6.08$^{+3.25}_{-2.46}$ & 0.35 & 0.33 & 0.02 & 40.78/42  \\
\hline
PCA & 96420-01-02-00 & 04-03-11 & SIMS/HIMS & 2.39$^{+0.05}_{-0.06}$ & 1.23$^{+0.03}_{-0.03}$ & 12.30$^{+1.93}_{-1.95}$ & 1.81 & 1.48 & 0.33 & 31.89/42  \\
\hline
PCA & 96420-01-02-01 & 06-03-11 & HIMS & 2.44$^{+0.05}_{-0.06}$ & 1.15$^{+0.16}_{-0.12}$ & 16.30$^{+5.35}_{-8.36}$ & 1.83 & 1.50 & 0.33 & 33.58/42  \\
\hline
PCA & 96420-01-02-02 & 08-03-11 & HIMS & 2.34$^{+0.04}_{-0.03}$ & 1.24$^{+0.02}_{-0.03}$ & 13.87$^{+1.35}_{-1.34}$ & 1.77 & 1.46 & 0.31 & 40.75/42  \\
\hline
PCA & 96420-01-02-03 & 10-03-11 & HIMS & 2.39$^{+0.05}_{-0.06}$ & 1.23$^{+0.03}_{-0.04}$ & 10.97$^{+1.92}_{-1.93}$ & 1.72 & 1.42 & 0.29 & 41.17/42 \\
\hline
PCA & 96420-01-03-00 & 12-03-11 & SIMS & 2.47$^{+0.03}_{-0.06}$ & 0.89$^{+0.08}_{-0.07}$ & 4.47$^{+1.15}_{-0.85}$ & 0.31 & 0.28 & 0.03 & 42.12/42 \\
\hline
PCA & 96420-01-05-02 & 25-03-11 & SIMS & 2.34$^{+0.02}_{-0.05}$ & 1.15$^{+0.04}_{-0.06}$ & 20.6$^{+2.11}_{-1.57}$ & 1.75 & 1.23 & 0.51 & 43.78/42 \\
\hline
\end{tabular}}
\end{center}
\tablecomments {$\Gamma$ is the power-law index, $kT_{in}$ is temperature of the inner disk edge in units of keV and N$_{diskbb}$ is the disk blackbody model normalization. Flux$_{total}$ is the total flux, Flux$_{diskbb}$ is the flux of the disk component and Flux$_{powerlaw}$ is the flux of the power-law component. All the fluxes are calculated in the energy range 2.0-60.0 keV for combined PCA and {\it SWIFT}/XRT spectra and 0.3-10.0 keV for {\it SWIFT}/XRT spectra only. Fluxes are given in a unit of 10$^{-9}$ ergs/s/cm$^2$.}
\end{table*}

\begin{table*}
\centering
 \caption{Summary of Comparative study of some characteristic parameters of IGR J17091$-$3624 and GRS 1915+105.} 
\begin{center}
\scalebox{0.75}{%
\begin{tabular}{|l|l|l|l|l|}
 \hline
\hline
 Characteristic & IGR J17091$-$3624 & GRS 1915+105  \\
 parameters & (variable state/$\rho$-like class) & (`$\rho$' class) \\
\hline
Oscillation timescale & $\sim$ 20 s $-$ 50 s & $\sim$ 40 s $-$ 110 s\\
\hline
Change in timescale & increases with time & increases with time \\
 & $\sim$ 1.32 mHz/day & $\sim$ 0.36 mHz/day \\
\hline
Peak to minimum & $\sim$ 2.91 $-$ 3.82 & $\sim$ 3.38 $-$ 4.28 \\
flux ratio & & \\
\hline
X-ray flux and & strong anti-correlation found & strong anti-correlation found \\
hard color relation & & \\
\hline
Variability pattern & slow exponential rise (t$_i$=t$_{rise}$ $\sim$ 14.67) & slow exponential rise (t$_i$=t$_{rise}$ $\sim$ 13.68) \\
(fitting with f(t)=ae$^{t/t_i}$) & fast decay (t$_i$=t$_{fall}$ $\sim$ -3.62) & fast decay (t$_i$=t$_{fall}$ $\sim$ -2.48) \\
\hline
QPOs & {26.3 mHz, 52.8 mHz} (on 31 Mar. 11) & {9.7 mHz \& 19.2 mHz} (on 26 May 97) \\
& {36.6 mHz, 75.9 mHz} (on 03 Apr. 11) & {12.1 mHz \& 24.3 mHz} (on 03 Jun. 97) \\
& {45.9 mHz, 94.2 mHz} (on 10 Apr. 11) etc. & {18.5 mHz, 37.9 mHz} (on 22 Jun. 97) \\
\hline
\end{tabular}}
\end{center}
\end{table*}
\clearpage

\begin{figure*}
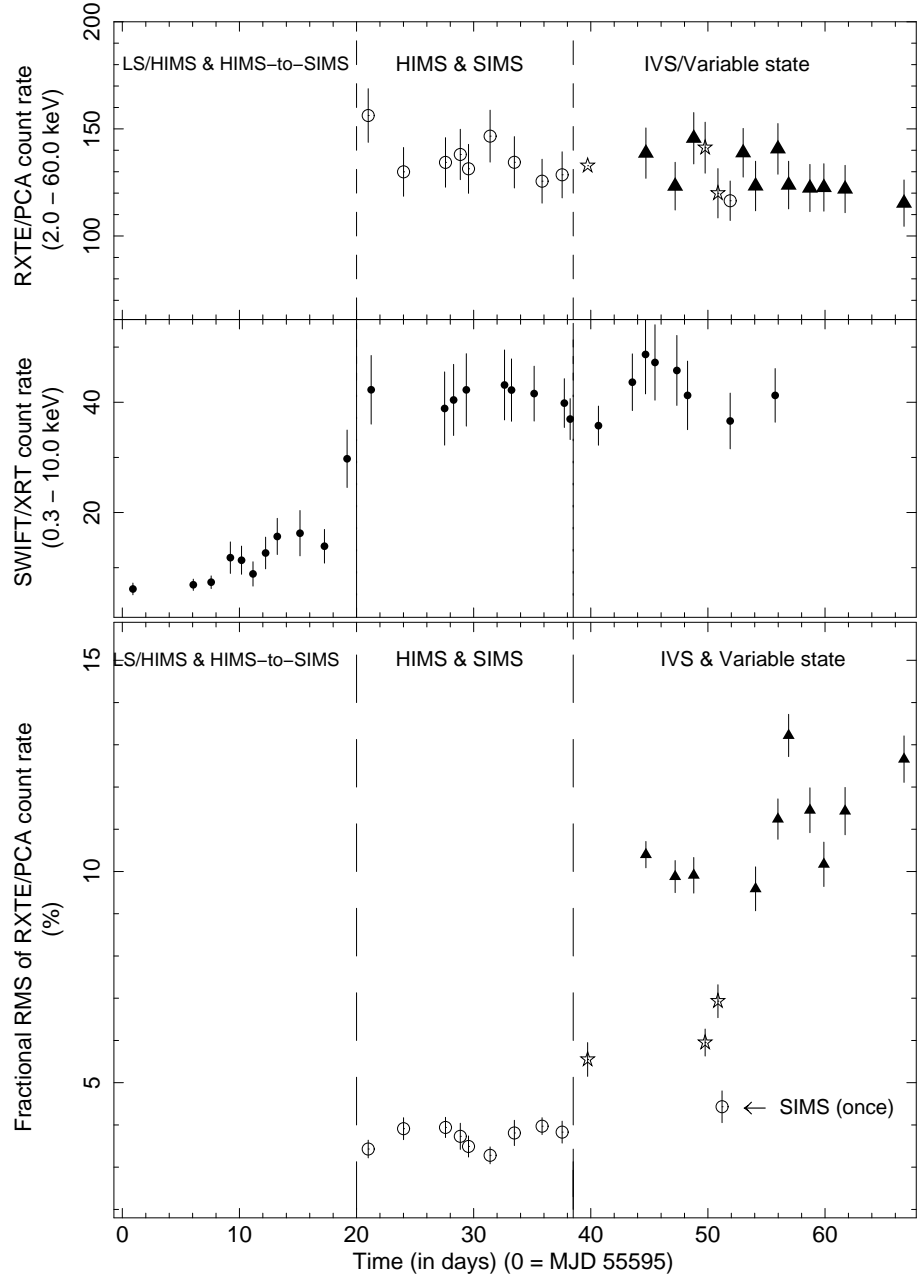

  \begin{center}
\includegraphics[scale=0.5,angle=-90]{fig1a.ps}
\includegraphics[scale=0.5,angle=-90]{fig1b1.ps}
\caption{Overall nature of 2011 outburst of IGRigr-paper.ps J17091$-$3624. {\it Top panel:} plot of background-subtracted {\it RXTE}/PCA average count rate between 2.0-60.0 keV with time, {\it Middle panel:} simultaneous plot of background-subtracted {\it SWIFT}/XRT average count rate between 0.3-10.0 keV with time, {\it Bottom panel:} fractional RMS values of RXTE/PCA count rate as observed during LS, HIMS, SIMS and variable state/$\rho$-like class. Due to contamination, {\it RXTE}/PCA has no data in first 20 days. HIMS \& SIMS, Intermediate variable state and variable state observations are shown by open circles, stars and triangles respectively in the top and bottom panels. Vertical lines separate the LS/HIMS and the HIMS \& SIMS and the variable state/$\rho$-like class. }
\end{center}
\end{figure*}
\clearpage

\begin{figure*}
  \begin{center}
\includegraphics[scale=10.0,angle=0]{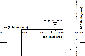}
\caption{0.3-10.0 keV filtered image of IGR J17091$-$3624 taken by the {\it SWIFT}/XRT in the photon counting mode on 03 February 11.}
\end{center}
\end{figure*}
\clearpage

\begin{figure*}
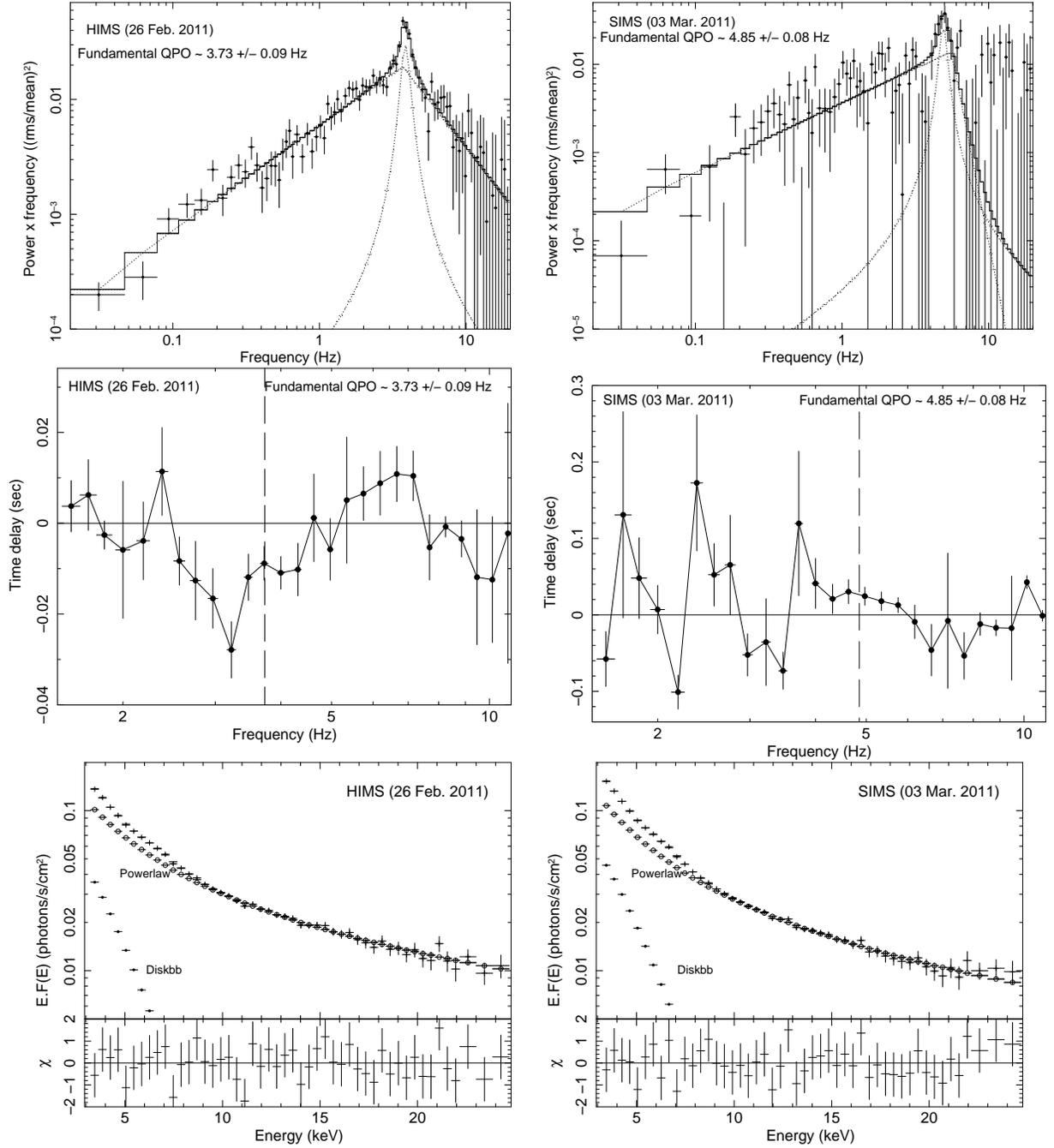

  \begin{center}
\includegraphics[scale=0.32,angle=-90]{fig3a.ps}
\includegraphics[scale=0.32,angle=-90]{fig3b.ps}
\includegraphics[scale=0.33,angle=-90]{fig3c.ps}
\includegraphics[scale=0.33,angle=-90]{fig3d.ps}
\includegraphics[scale=0.33,angle=-90]{fig3e.ps}
\includegraphics[scale=0.33,angle=-90]{fig3f.ps}
\caption{{\it SIMS and HIMS in IGR J17091$-$3624 as observed by {\it RXTE}/PCA}. Top panel : Power density spectra fitted with broken powerlaw and Lorentzians are shown during the hard intermediate state (HIMS) as observed on 26 February, 2011 (top left panel) and during soft intermediate state (SIMS) as observed on 03 March, 2011 (top right panel). Middle panel : frequency dependent time-lag spectra, calculated between 1.8$-$4.2 keV and 5.0$-$13.0 keV energy bands, are shown for same observations during the HIMS (middle left panel) and the SIMS (middle right panel). Dotted vertical lines show the fundamental QPO frequency in both panels. Bottom panels : {\it RXTE}/PCA spectra fitted with {\tt diskbb+powerlaw} in the range 3.0-25.0 keV are shown for the HIMS (bottom left panel) and the SIMS (bottom right panel) along with their fitted model components and residuals.}
\end{center}
\end{figure*}
\clearpage

\begin{figure*}
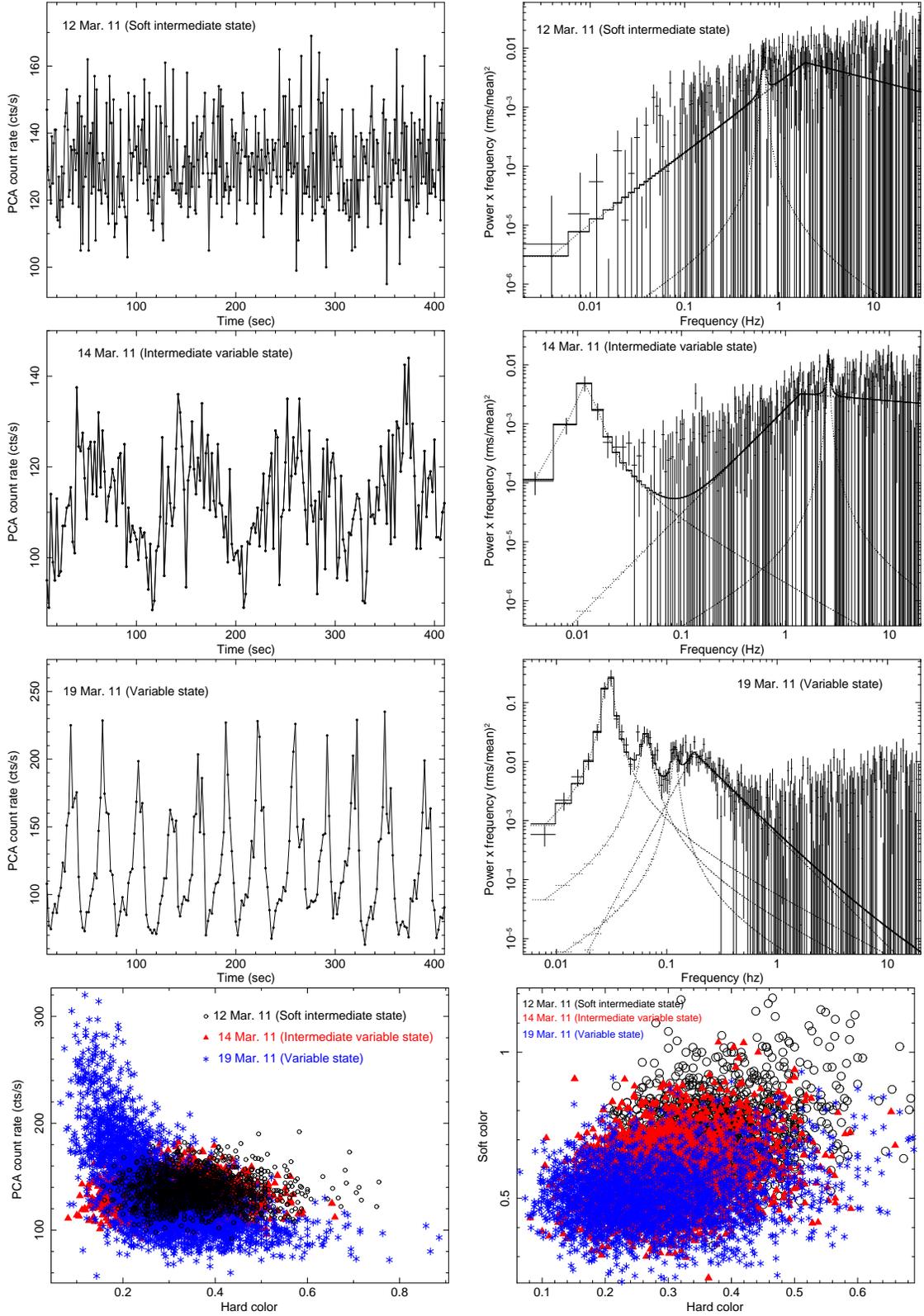

  \begin{center}
\includegraphics[scale=0.30,angle=-90]{fig4a.ps}
\includegraphics[scale=0.30,angle=-90]{fig4b.ps}
\includegraphics[scale=0.30,angle=-90]{fig4c.ps}
\includegraphics[scale=0.30,angle=-90]{fig4d.ps}
\includegraphics[scale=0.30,angle=-90]{fig4e.ps}
\includegraphics[scale=0.30,angle=-90]{fig4f.ps}
\includegraphics[scale=0.30,angle=-90]{fig4g.ps}
\includegraphics[scale=0.30,angle=-90]{fig4h.ps}
\caption{{\it Transition from low-variability class (soft intermediate state) to high-variability class (variable state) via intermediate in IGR J17091$-$3624}. Top three left panels show 2.0-60.0 keV {\it RXTE}/PCA light curves of the SIMS, IVS and Variable state respectively. Top three right panels show 2.0-60.0 keV {\it RXTE}/PCA rms normalized power density spectra of the SIMS, IVS and Variable state respectively, fitted with broken power-law and Lorentzian. {\it Bottom left and right panel:} evolution of classes from the SIMS to the variable state/$\rho$-like class in the Hardness Intensity Diagram (HID) and the color color Diagram (CD) respectively. Each point in the HID and CD are created using 1 sec time binning for the entire observation and errors are $\le$ 7\%.}
\end{center}
\end{figure*}
\clearpage

\begin{figure*}
  \begin{center}
\includegraphics[scale=0.62,angle=-90]{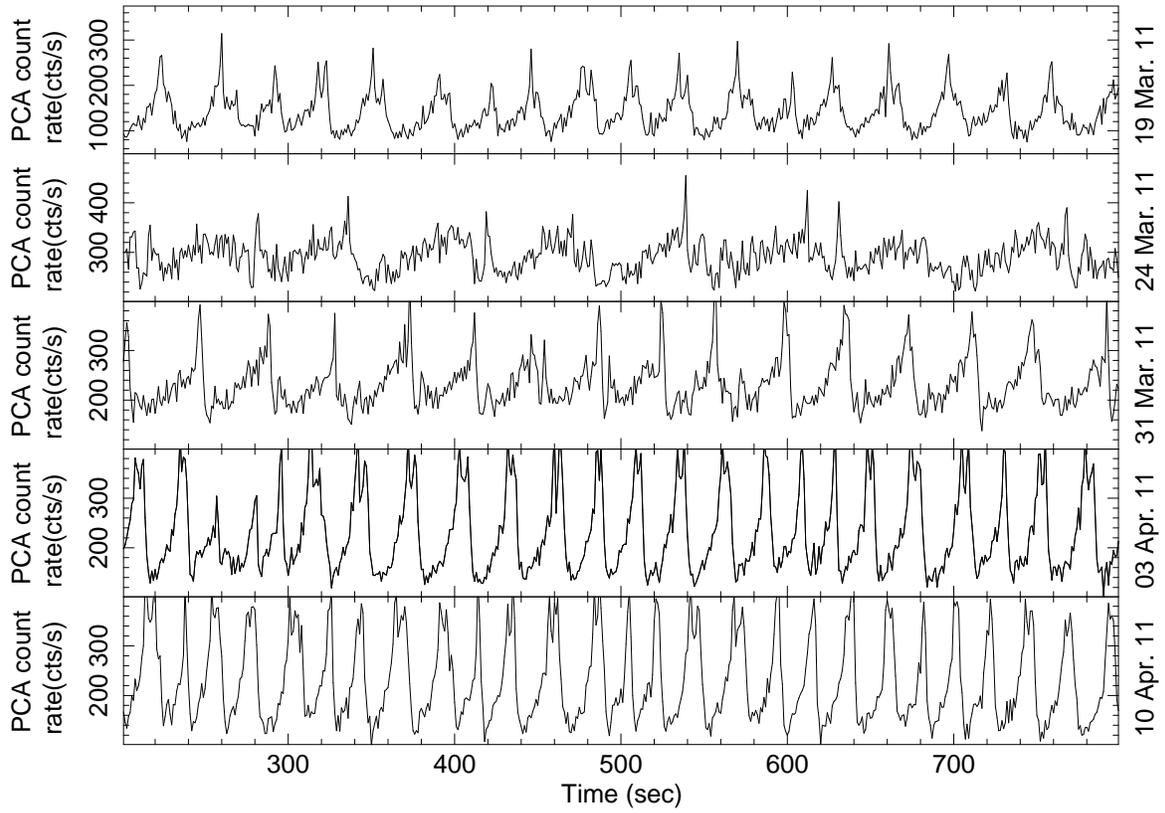}
\caption{Lightcurve evolution of the variability pattern during variable state/$\rho$-like class in IGR J17091$-$3624. All 5 panels show 2.0-60.0 keV background-subtracted PCA light curve for five different observations from 19 March 11 to 10 April 11 with 1s bintime.}
\end{center}
\end{figure*}
\clearpage

\begin{figure*}
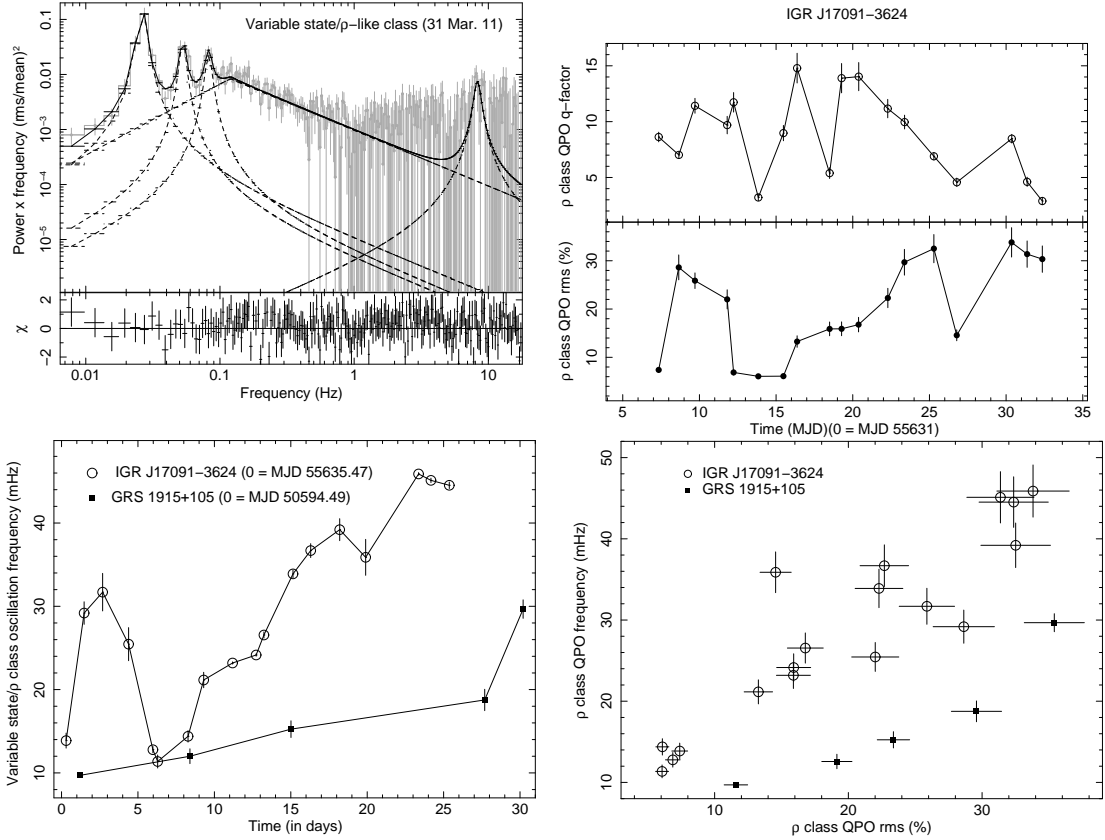

  \begin{center}
\includegraphics[scale=0.30,angle=-90]{fig6a.ps}
\includegraphics[scale=0.30,angle=-90]{fig6b1.ps}
\includegraphics[scale=0.30,angle=-90]{fig6c1.ps}
\includegraphics[scale=0.30,angle=-90]{fig6d1.ps}
\caption{Power density spectral evolution during the variable state/$\rho$-like class in IGR J17091$-$3624. Top left panel : Example of the PDS of a variability state/$\rho$-like class, fitted with broken powerlaw and multiple Lorentzians (see Table 1 for parameter values) is shown as observed on 31 March, 2011. Fitted model components are shown in dotted lines alongwith the residuals. Top right panel :  the time evolution of the quality factor and the rms amplitude(\%) of the characteristic pulsation at the fundamental frequency during variable state/$\rho$-like class observations are shown in the upper and lower top right panels respectively. Bottom left panel : the time evolution of the characteristic oscillation frequencies in the variable state/$\rho$-like class observations from both IGR J17091$-$3624 (hollow circles) and GRS 1915+105 (solid squares) are shown respectively. Bottom right panel : the plot of characteristic oscillation frequencies in the variable state/$\rho$-like class observations as a function of the rms amplitude(\%) of the characteristic pulsation is shown for both IGR J17091-3624 (hollow circles) and GRS 1915+105 (solid squares).}
\end{center}
\end{figure*}
\clearpage

\begin{figure*}
  \begin{center}
\includegraphics[scale=0.50,angle=-90]{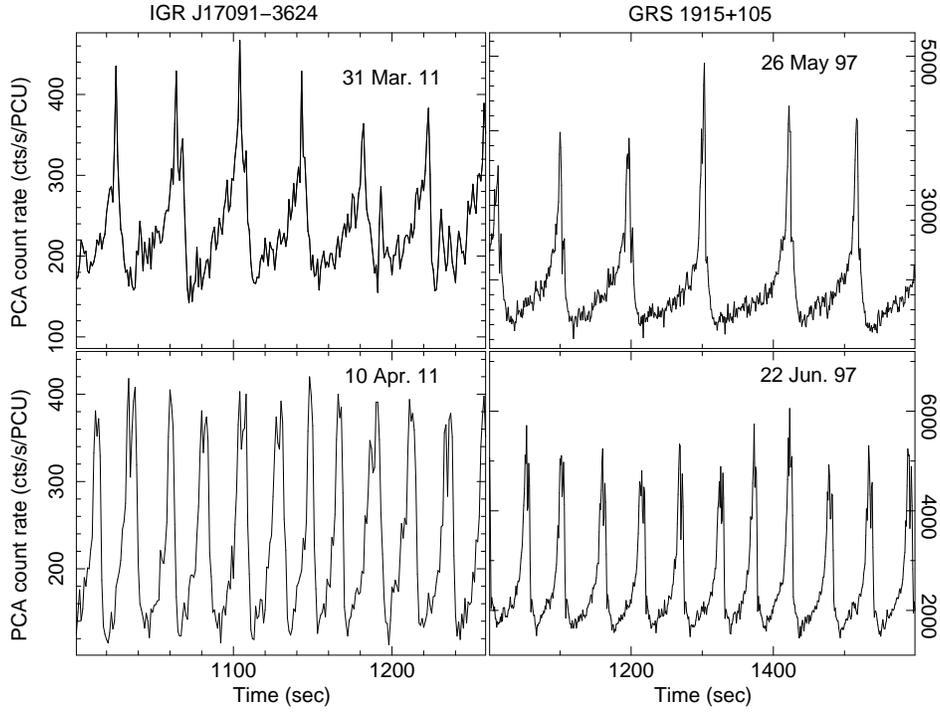}
\caption{Comparative study between the variable state/$\rho$-like class in IGR J17091$-$3624 and the `$\rho$' class of GRS 1915+105. Left panel figures correspond to 2.0-60.0 keV {\it RXTE}/PCA light curve of IGR J17091$-$3624 on 31 March 11 (top) and 10 April 11 (bottom) and the right panel figures corresponds to 2.0-60.0 keV {\it RXTE}/PCA light curve of GRS 1915+105 on 26 May 97 (top) and 22 June 97 (bottom). A decrease in variability time scale with time is common to both sources as they evolve.}
\end{center}
\end{figure*}

\begin{figure*}
  \begin{center}
\includegraphics[scale=0.50,angle=-90]{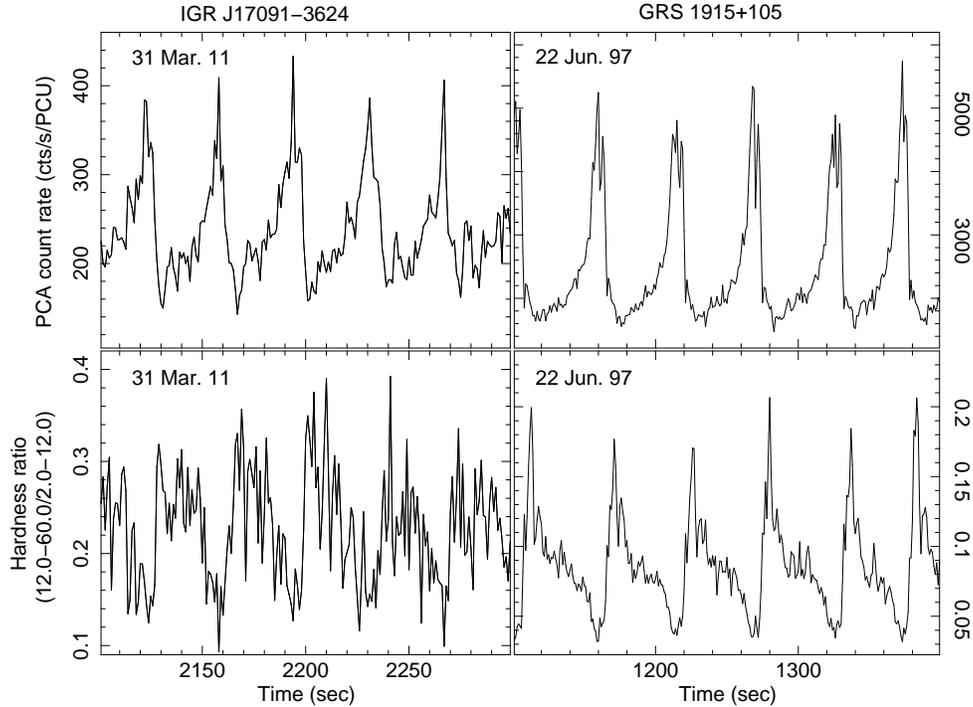}
\caption{Comparative study between the variable state/$\rho$-like class variability class of IGR J17091$-$3624 and the `$\rho$' class of GRS 1915+105. Top panels show the 2.0-60.0 keV {\it RXTE}/PCA light curve of IGR J17091$-$3624 (left) and GRS 1915+105 (right). Bottom panels show the plot of hardness ratio (defined as the ratio of {\it RXTE}/PCA count rate between 12.0-60.0 keV and 2.0-12.0 keV) with time for IGR J17091$-$3624 (left) \& GRS 1915+105 (right). Strong anti correlation between X-ray flux and hardness ratio is found in both sources.}
\end{center}
\end{figure*}
\clearpage

\begin{figure*}
  \begin{center}
\includegraphics[scale=0.30,angle=0]{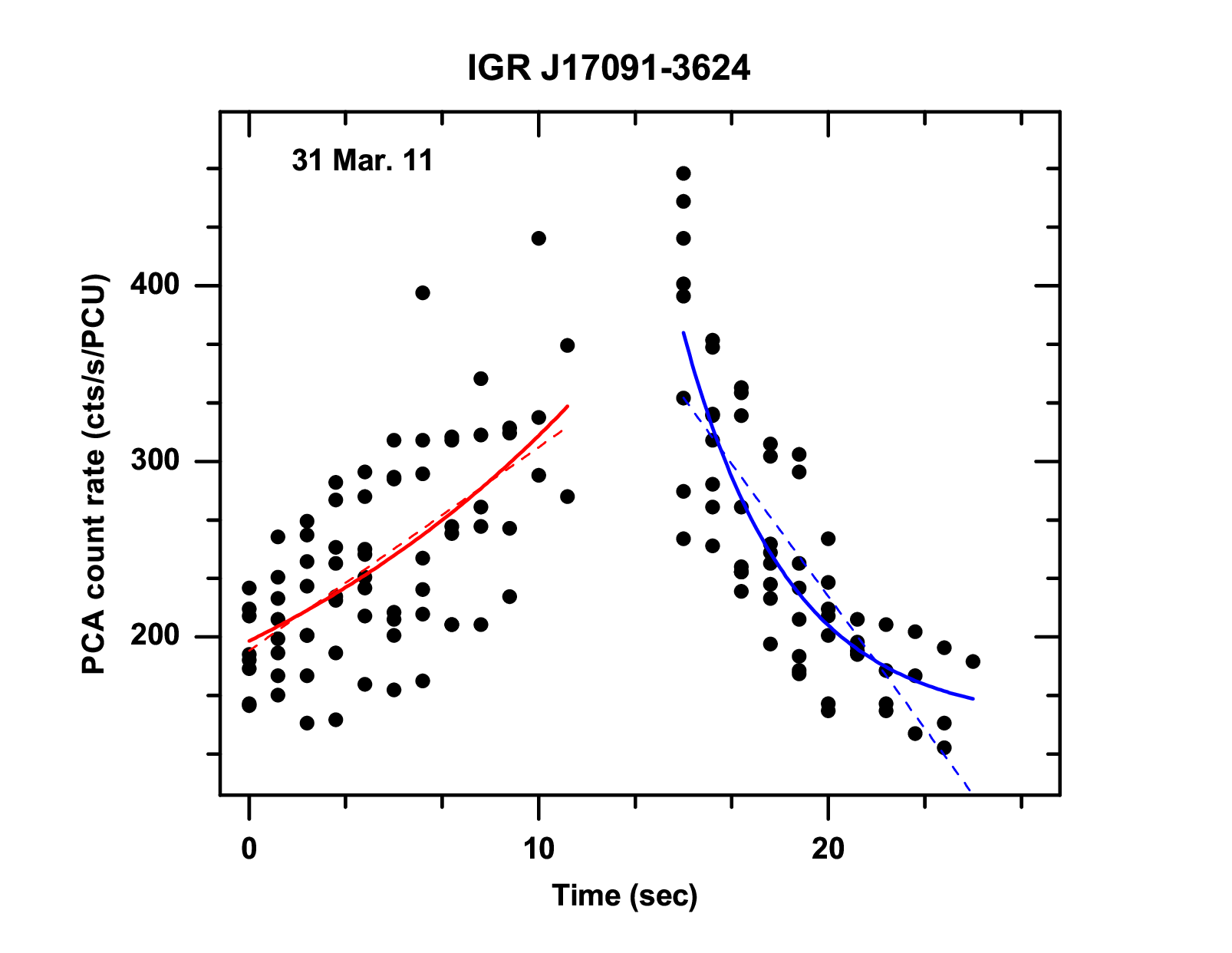}
\includegraphics[scale=0.30,angle=0]{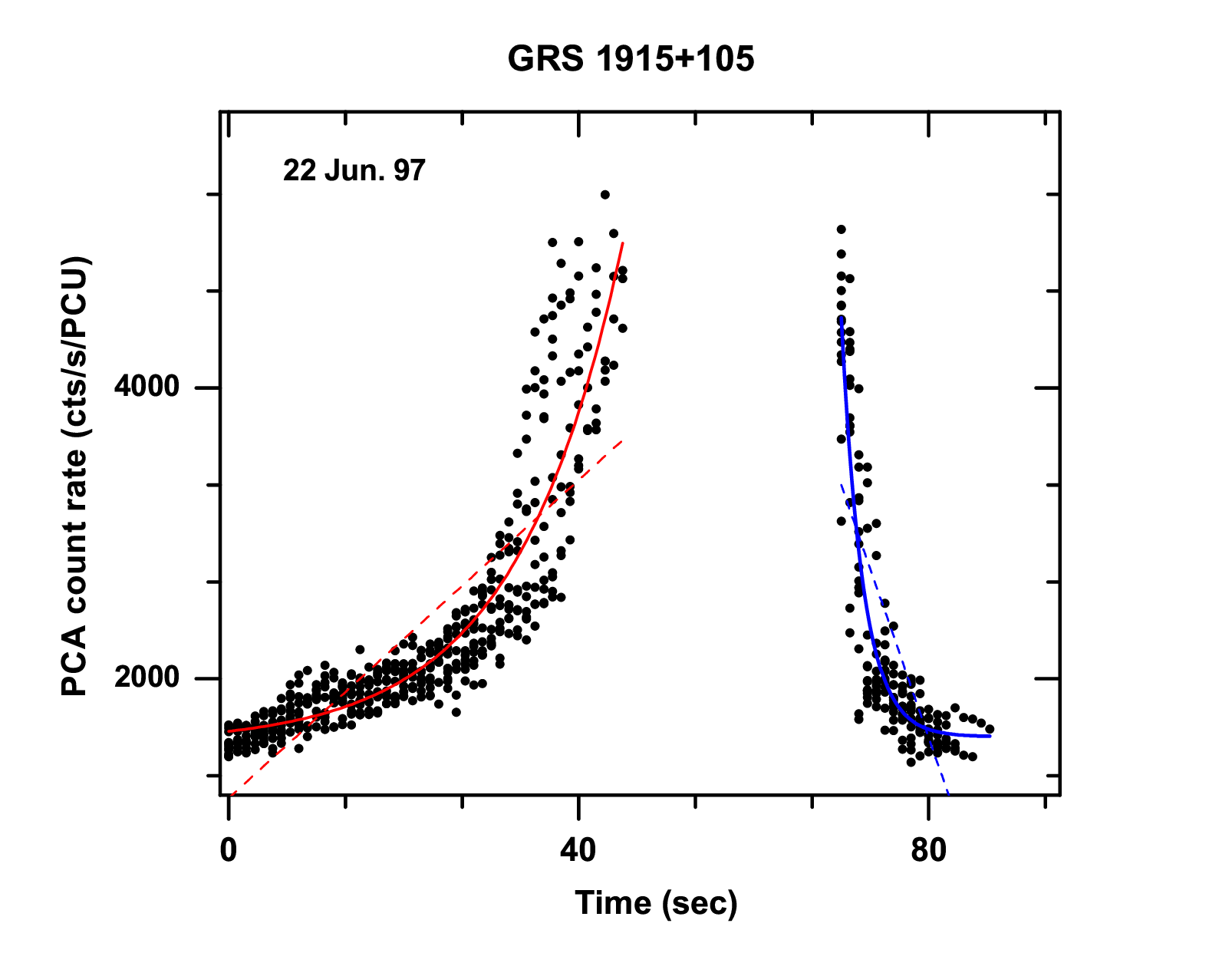}
\includegraphics[scale=0.305,angle=0]{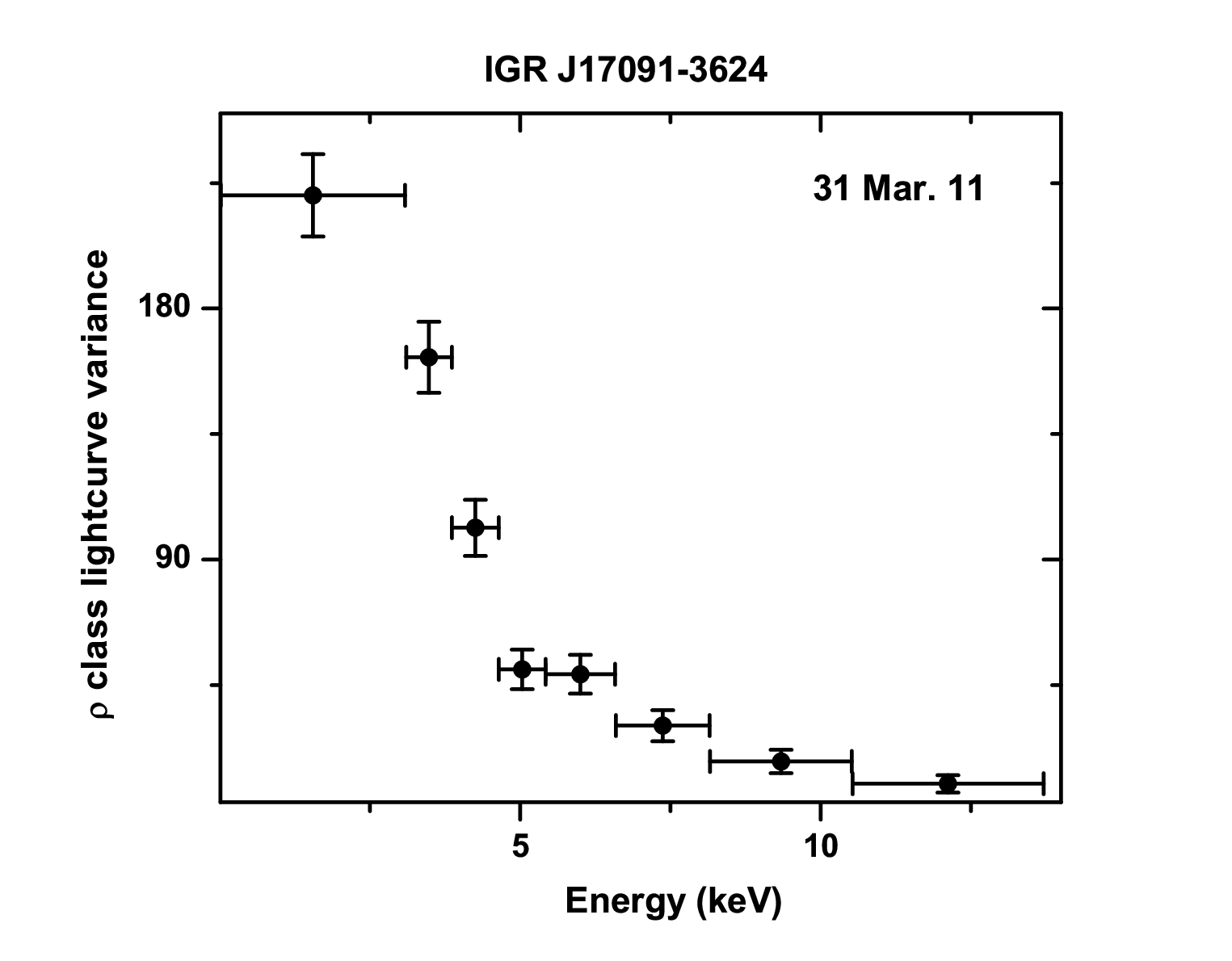}
\includegraphics[scale=0.305,angle=0]{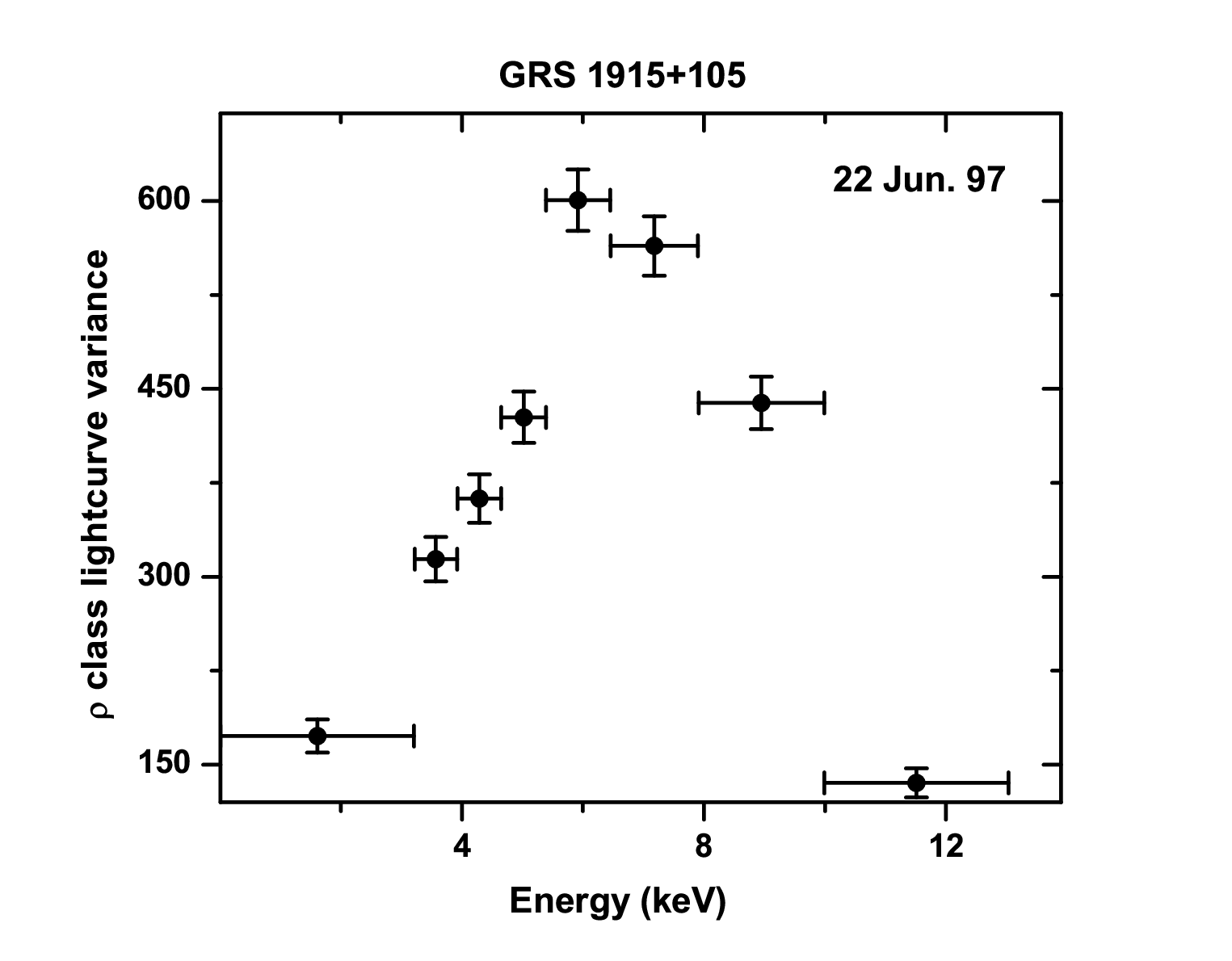}
\caption{Comparative study between the variable state/$\rho$-like class variability class of IGR J17091$-$3624 and the `$\rho$' class of GRS 1915+105. {\it Top panels:} combined rise profile (left) and decay profile (right) of several bursts in IGR J17091$-$3624 (top left panel) and GRS 1915+105 (top right panel). Start of all rise profiles are normalized to 0 sec in both sources and start of all decay profiles are normalized to 15 sec in IGR J17091-3624 and 70 sec in GRS 1915+105. In both sources, an exponential growth function (ae$^{x/t}$) and an exponential decay function (ae$^{-x/t}$) (shown by solid lines) can fit the rise and decay profiles respectively, better than a straight line (shown by dotted line) with the significance $>$3$\sigma$. Bottom panels show the lightcurve variance spectra during `heartbeat' oscillations in IGR J17091-3624 (bottom left panel) and GRS 1915+105 (bottom right panel). A discontinuity is observed in the spectra at $\sim$6 keV in GRS 1915+105 whereas the variance spectra of IGR J17091-3624 is monotonous.}
\end{center}
\end{figure*}
\clearpage

\begin{figure*}
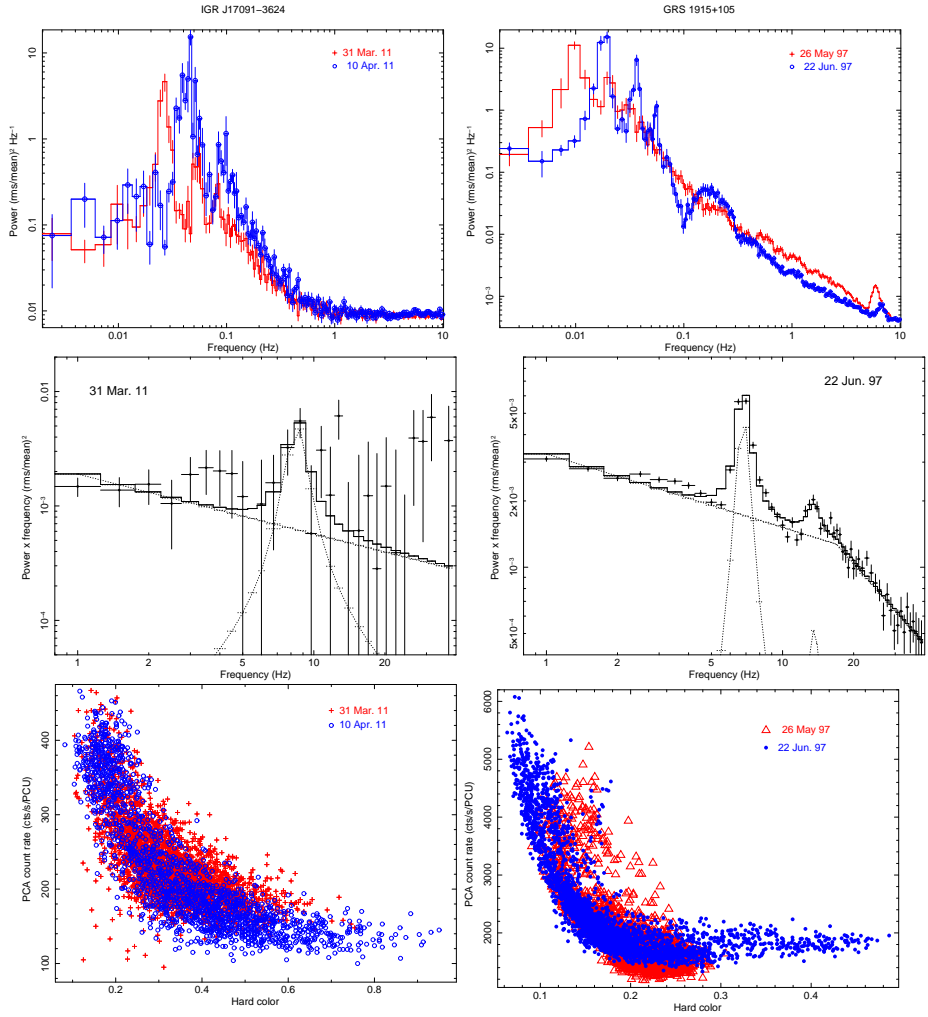

  \begin{center}
\includegraphics[scale=0.25,angle=-90]{fig10a.ps}
\includegraphics[scale=0.25,angle=-90]{fig10b.ps}
\includegraphics[scale=0.25,angle=-90]{fig10c.ps}
\includegraphics[scale=0.25,angle=-90]{fig10d.ps}
\includegraphics[scale=0.25,angle=-90]{fig10e.ps}
\includegraphics[scale=0.25,angle=-90]{fig10f.ps}
\caption{All left panel figures correspond to IGR J17091$-$3624 and all right panel figures correspond to GRS 1915+105. {\it Top panels:} 2.0-60.0 keV {\it RXTE}/PCA rms normalized power density spectra. {\it Middle panels:} Typical type-A 7$-$10 Hz QPO in the PDS of IGR J17091$-$3624 (middle left) and a type-B QPO in the PDS of GRS 1915+105 (middle right) during the variable state/$\rho$ class. {\it Lower  panels:} Hardness Intensity Diagrams for the same definition of hard color and intensity. Both panels show that both sources evolved in a similar way in the PDS as well as in HIDs as the characteristic oscillation frequency changes.}
\end{center}
\end{figure*}
\clearpage


\clearpage 

\begin{thebibliography}{99}
\bibitem[\protect\citeauthoryear{Altamirano et al.}{2013}]{b80} Altamirano, D., Wijnands, R., \& Belloni, T., 2013, Astron. Telegram, 5112, 1
\bibitem[\protect\citeauthoryear{Altamirano \& Strohmayer}{2012a}]{b54} Altamirano, D., Strohmayer, T., 2012a, ApJ, 754, L23
\bibitem[\protect\citeauthoryear{Altamirano \& Belloni}{2012b}]{b94} Altamirano, D. \& Belloni, T. 2012b, ApJ, 747, L4
\bibitem[\protect\citeauthoryear{Altamirano et al.}{2011a}]{b6} Altamirano, D., Linares, M., van der Klis, M., Wijnands, R., Kalamkar, M., Casella, P., Watts, A. \& Patruno, A. 2011a, Astron. Telegram, 3225, 1
\bibitem[\protect\citeauthoryear{Altamirano et al.}{2011b}]{b26} Altamirano, D., Belloni, T., Krimm, H. A., Casella, P., Curran, P., Kennea, J. A., Kalamkar, M. \& van der Klis, M. 2011b, Astron. Telegram, 3230, 1 
\bibitem[\protect\citeauthoryear{Altamirano et al.}{2011c}]{b95} Altamirano, D., Belloni, T., Linares, M., van der Klis, M., Wijnands, R., Curran, P. A., Kalamkar, M., Stiele, H., Motta, S., Muñoz-Darias, T., Casella, P., Krimm, H, 2011c, ApJ, 742, L17
\bibitem[\protect\citeauthoryear{Belloni et al.}{2005}]{b86} Belloni, T., Homan, J., Casella, P., van der Klis, M., Nespoli, E., Lewin, W. H. G., Miller, J. M., \& Me\'ndez, M. 2005, A\&A, 440, 207
\bibitem[\protect\citeauthoryear{Belloni et al.}{2000}]{b16} Belloni, T., Klein-Wolt, M., Me\'ndez, M., van der Klis, M. \& van Paradijs, J. 2000, A\&A, 355, 271
\bibitem[\protect\citeauthoryear{Bhattacharyya}{2010}]{b34} Bhattacharyya, S., 2010, AdSpR, 45, 949
\bibitem[\protect\citeauthoryear{Bodaghee et al.}{2012}]{b82} Bodaghee, A., Rahoui, F., Tomsick, J. A., Rodriguez, J., 2012, ApJ, 751, 113
\bibitem[\protect\citeauthoryear{Burrows et al.}{2005}]{b32} Burrows, David N., Hill, J. E., Nousek, J. A., Kennea, J. A., Wells, A., Osborne, J. P., Abbey, A. F., Beardmore, A. et al., 2005, Space Sc. Rev., 120, 165
\bibitem[\protect\citeauthoryear{Caballero-Garci\'a et al.}{2009}]{b43} Caballero-Garci\'a, M. D., Miller, J. M., Trigo, M. Di\'az, Kuulkers, E., Fabian, A. C., Mas-Hesse, J. M., Steeghs, D. \& van der Klis, M. 2009, ApJ, 692, 1339
\bibitem[\protect\citeauthoryear{Capitanio et al.}{2012}]{b98} Capitanio, F., Del Santo, M., Bozzo, E., Ferrigno, C., De Cesare, G., \& Paizis, A., 2012, MNRAS, 422, 3130
\bibitem[\protect\citeauthoryear{Capitanio et al.}{2011}]{b25} Capitanio, F., Tramacere, A., Del Santo, M., Bozzo, E., Watanabe, K., Caballero, I., Chenevez, J.,  Paizis, A. et al. 2011, Astron. Telegram, 3159, 1
\bibitem[\protect\citeauthoryear{Capitanio et al.}{2009}]{b3} Capitanio, F., Giroletti, M., Molina, M., Bazzano, A., Tarana, A., Kennea, J., Dean, A. J., Hill, A. B. et al. 2009, ApJ, 690, 1621
\bibitem[\protect\citeauthoryear{Capitanio et al.}{2006}]{b1} Capitanio, F., Bazzano, A., Ubertini, P., Zdziarski, A. A., Bird, A. J., De Cesare, G., Dean, A. J., Stephen, J. B. \& Tarana, A. 2006, ApJ, 643, 376
\bibitem[\protect\citeauthoryear{Casella et al.}{2005}]{b81} Casella, P., Belloni, T. \& Stella, L., 2005, ApJ, 629, 403
\bibitem[\protect\citeauthoryear{Castro-Tirado et al.}{1994}]{b14} Castro-Tirado, A. J., Brandt, S., Lund, N., Lapshov, I., Sunyaev, R. A., Shlyapnikov, A. A., Guziy, S. \& Pavlenko, E. P. 1994, ApJS, 92, 469
\bibitem[\protect\citeauthoryear{Chaty et al.}{2008}]{b24} Chaty, S., Rahoui, F., Foellmi, C., Tomsick, J. A., Rodriguez, J. \& Walter, R. 2008, A\&A, 484, 783
\bibitem[\protect\citeauthoryear{Del Santo et al.}{2009}]{b37} Del Santo, M., Belloni, T. M., Homan, J., Bazzano, A., Casella, P., Fender, R. P., Gallo, E., Gehrels, N. et al. 2009, MNRAS, 392, 992 
\bibitem[\protect\citeauthoryear{Del Santo et al.}{2011}]{b8} Del Santo, M., Kuulkers, E.; Bozzo, E., Capitanio, F., Alfonso-Garzon, J., Beckmann, V., Bird, T., Brandt, S. et al. 2011, Astron. Telegram, 3203, 1
\bibitem[\protect\citeauthoryear{Dunn et al.}{2008}]{b38} Dunn, R. J. H., Fender, R. P., K{\"o}rding, E. G., Cabanac, C. \& Belloni, T. 2008, MNRAS, 387, 545
\bibitem[\protect\citeauthoryear{Eikenberry et al.}{1998}]{b18} Eikenberry, S. S., Matthews, K., Murphy, T. W., Jr., Nelson, R. W., Morgan, E. H., Remillard, R. A. \& Muno, M. 1998, ApJ, 506, L31
\bibitem[\protect\citeauthoryear{Fender, Belloni \& Gallo}{2004}]{b83} Fender, R. P., Belloni, T. M., Gallo, E., 2004, MNRAS, 355, 1105
\bibitem[\protect\citeauthoryear{Frank, King \& Raine}{2002}]{b23} Frank, J., King, A., Raine, D., 2002, {\it Accretion Power in Astrophysics}, 3rd Edition, Cambridge University Press
\bibitem[\protect\citeauthoryear{Fujimoto et al.}{1981}]{b53} Fujimoto, M. Y., Hanawa, T., \& Miyaji, S., 1981, ApJ, 247, 267
\bibitem[\protect\citeauthoryear{Greiner}{1994}]{b20} Greiner, J., 1994, AGAb, 10, 23
\bibitem[\protect\citeauthoryear{Homan \& Belloni}{2005}]{b90} Homan, J., Belloni, T., 2005, Ap\&SS, 300, 107
\bibitem[\protect\citeauthoryear{In't Zand et al.}{2003}]{b88} In't Zand, J. J. M., Heise, J., Lowes, P., \& Ubertini, P. 2003, Astron. Telegram, 160, 1
\bibitem[\protect\citeauthoryear{Jahoda et al.}{1996}]{b27} Jahoda, K., Swank, J. H., Giles, A. B., Stark, M. J., Strohmayer, T., Zhang, WT., \& Morgan, E. H. 1996, SPIE, 2808, 59
\bibitem[\protect\citeauthoryear{Janiuk \& Misra}{2012}]{b55} Janiuk, A., Misra, R., 2012, A\&A, 540, 114
\bibitem[\protect\citeauthoryear{Joinet et al.}{2007}]{b41} Joinet, A., Jourdain, E., Malzac, J., Roques, J. P., Corbel, S., Rodriguez, J. \& Kalemci, E. 2007, ApJ, 657, 400
\bibitem[\protect\citeauthoryear{Klein-Wolt et al.}{2002}]{b35} Klein-wolt, M., et al., 2002, MNRAS, 331, 745
\bibitem[\protect\citeauthoryear{Krimm \& Kennea}{2011b}]{b10}  Krimm, H. A. \& Kennea, J. A. 2011, Astron. Telegram, 3148, 1
\bibitem[\protect\citeauthoryear{Krimm et al.}{2011a}]{b11} Krimm, H. A., Barthelmy, S. D., Baumgartner, W., Cummings, J., Fenimore, E., Gehrels, N., Kennea, J. A., Markwardt, C. B. et al. 2011, Astron. Telegram, 3144, 1
\bibitem[\protect\citeauthoryear{King et al.}{2012}]{b96} King, A. L., Miller, J. M., Raymond, J., Fabian, A. C., Reynolds, C. S., Kallman, T. R., Maitra, D., Cackett, E. M., Rupen, M. P., 2012, ApJL, 746, 20 
\bibitem[\protect\citeauthoryear{Kuulkers et al.}{2003}]{b19} Kuulkers, E., Lutovinov, A., Parmar, A., Capitanio, F., Mowlavi, N. \& Hermsen, W. 2003, Astron. Telegram, 149, 1
\bibitem[\protect\citeauthoryear{Kubota et al.}{1998}]{b84} Kubota, A., Tanaka, Y., Makishima, K., Ueda, Y., Dotani, T., Inoue, H., Yamaoka, K., 1998, PASJ, 50, 667
\bibitem[\protect\citeauthoryear{Lewin et al.}{1995}]{b52} Lewin, W. H. G., van Paradijs, J., \& Taam, R. E. 1995, in X-Ray Binaries, ed. W. H. G. Lewin, J. van Paradijs, \& E. P. J. van den Heuvel (Cambridge: Cambridge Univ. Press), 175
\bibitem[\protect\citeauthoryear{Lochner et al.}{1992}]{b21} Lochner, J., Whitlock, L., Kouveliotou, C., 1992, IAUC, 5658, 2
\bibitem[\protect\citeauthoryear{Lutovinov et al.}{2005}]{b31} Lutovinov, A. A., Revnivtsev, M., Molkov, S. \& Sunyaev, R. 2005, A\&A, 430, 997
\bibitem[\protect\citeauthoryear{Lutovinov et al.}{2003}]{b22} Lutovinov, A. A. \& Revnivtsev, M. G. 2003, Astron. Lett., 29, 719
\bibitem[\protect\citeauthoryear{Markoff et al.}{2005}]{b48} Markoff, S., Nowak, M. A., \& Wilms, J. 2005, ApJ, 635, 1203
\bibitem[\protect\citeauthoryear{Massaro et al.}{2010}]{b4}  Massaro, E., Ventura, G., Massa, F., Feroci, M., Mineo, T., Cusumano, G., Casella, P. \& Belloni, T. 2010, A\&A, 513, 21
\bibitem[\protect\citeauthoryear{Miller}{2007}]{b42} Miller, J. M. 2007, ARA\&A, 45, 441
\bibitem[\protect\citeauthoryear{Mineo et al.}{2006}]{b33} Mineo, T., Romano, P., Mangano, V., Moretti, A., Cusumano, G., La Parola, V., Troja, E., Campana, S. et al., 2006, Nuovo Cimento B, 121, 1521
\bibitem[\protect\citeauthoryear{Mirabel et al.}{1998}]{b97} Mirabel, I. F., Dhawan, V., Chaty, S., Rodriguez, L. F., Marti, J., Robinson, C. R., Swank, J., \& Geballe, T., 1998, A\&A, 330, L9
\bibitem[\protect\citeauthoryear{Mirabel \& Rodriguez}{1994}]{b15} Mirabel, I. F.\& Rodriguez, L. F. 1994, Nature, 371, 46
\bibitem[\protect\citeauthoryear{Muno et al.}{1999}]{b89} Muno, M. P., Morgan, E. H. \& Remillard, R. A., 1999, ApJ, 527, 321
\bibitem[\protect\citeauthoryear{Mun\~oz-Darias et al.}{2008}]{b39} Mun\~oz-Darias, T., Casares, J., \& Marti\'nez-Pais, I. G. 2008, MNRAS, 385, 2205
\bibitem[\protect\citeauthoryear{Neilsen et al.}{2012}]{b51} Neilsen, J., Remillard, R. A., Lee, J. C., 2012, ApJ, 750, 71
\bibitem[\protect\citeauthoryear{Pahari et al.}{2013a}]{b61} Pahari, M., et al., 2013a, ApJ, 778, 46
\bibitem[\protect\citeauthoryear{Pahari et al.}{2013b}]{b62} Pahari, M., et al., 2013b, MNRAS, 436, 2334
\bibitem[\protect\citeauthoryear{Pahari et al.}{2013c}]{b63} Pahari, M., et al., 2013c, ApJ, 778, 136
\bibitem[\protect\citeauthoryear{Pahari et al.}{2011a}]{b28} Pahari, M., Bhattacharyya, S., Yadav, J. S. \& Pandey, S. K., 2011, MNRAS, 422, L87
\bibitem[\protect\citeauthoryear{Pahari et al.}{2011b}]{b50} Pahari, M., Bhattacharyya, S. \& Yadav, J. S. 2011, Astron. Telegram, 3266, 1
\bibitem[\protect\citeauthoryear{Pahari \& Pal}{2010}]{b44} Pahari, M. \& Pal, S. 2010, MNRAS, 409, 903
\bibitem[\protect\citeauthoryear{Paul et al.}{1998}]{b40} Paul, B., Agrawal, P. C., Rao, A. R., Vahia, M. N., Yadav, J. S., Marar, T. M. K., Seetha, S. \& Kasturirangan, K. 1998, A\&AS, 128, 145
\bibitem[\protect\citeauthoryear{Rao et al.}{2000}]{b99} Rao, A. R., Yadav, J. S. \& Paul, B., 2000, ApJ, 544, 443
\bibitem[\protect\citeauthoryear{Rebusco et al.}{2012}]{b85} Rebusco, P., Moskalik, P., Kluz\'niak, W., Abramowicz, M. A., 2012, A\&A, 540, L4
\bibitem[\protect\citeauthoryear{Reig et al.}{2000}]{b87} Reig, P., Belloni, T., van der Klis, M., Me\'ndez, M., Kylafis, N. D., \& Ford, E. C., 2000, ApJ, 541, 883
\bibitem[\protect\citeauthoryear{Remillard \& McClintock}{2006}]{b93} Remillard, R. A. \& McClintock, J. E., 2006, ARA\&A, 44, 49
\bibitem[\protect\citeauthoryear{Rodriguez et al.}{2011a}]{b13} Rodriguez, J., Corbel, S., Tomsick, J. A., Paizis, A. \& Kuulkers, E. 2011a, Astron. Telegram, 3168, 1
F\bibitem[\protect\citeauthoryear{Rodriguez et al.}{2011b}]{b92} Rodriguez, J., Corbel, S., Caballero, I., Tomsick, J. A., Tzioumis, T., Paizis, A., Cadolle, Bel M., \& Kuulkers, E., 2011b, A\&A, 533, L4
\bibitem[\protect\citeauthoryear{Rothschild et al.}{1998}]{b106} Rothschild, R. E., Blanco, P. R., Gruber, D. E., Heindl, W. A., MacDonald, D. R., Marsden, D. C., Pelling, M. R., Wayne, L. R. \& Hink, P. L. 1998, ApJ, 496, 538
\bibitem[\protect\citeauthoryear{Shaposhnikov}{2011}]{b12} Shaposhnikov, Nikolai 2011, Astron. Telegram, 3179, 1
\bibitem[\protect\citeauthoryear{Sheskin}{2003}]{b56} Sheskin, D. J., 2003, Handbook of Parametric and Nonparametric Statistical Procedures: Third Edition, CRC Press
\bibitem[\protect\citeauthoryear{Soleri et al.}{2008}]{b30} Soleri, P., Belloni, T. \& Casella, P. 2008, MNRAS, 383, 1089
\bibitem[\protect\citeauthoryear{Steeghs et al.}{2013}]{b79} Steeghs, D., McClintock, J. E., Parsons, S. G., et al., 2013, ApJ, 768, 185
\bibitem[\protect\citeauthoryear{Sunyaev \& Revnivtsev}{2000}]{b91} Sunyaev, R. \& Revnivtsev, M., 2000, A\&A, 358, 617 
\bibitem[\protect\citeauthoryear{Taam et al.}{1997}]{b2} Taam, R. E., Xingming, C. \& Swank, J., E. 1997, APJ, 485, L83
\bibitem[\protect\citeauthoryear{Tingay et al.}{1995}]{b17} Tingay, S. J., Jauncey, D. L., Preston, R. A., Reynolds, J. E., Meier, D. L., Murphy, D. W., Tzioumis, A. K., McKay, D. J. et al. 1995, Nature, 374, 141
\bibitem[\protect\citeauthoryear{Yadav et al.}{1999}]{b5} Yadav, J. S., Rao, A. R., Agrawal, P. C., Paul, B., Seetha, S. \& Kasturirangan, K. 1999, ApJ, 517, 935
\bibitem[\protect\citeauthoryear{Yadav}{2001}]{b66} Yadav, J. S., 2001, ApJ, 548, 876
\bibitem[\protect\citeauthoryear{Yadav}{2006}]{b67} Yadav, J. S., 2006, ApJ, 646, 385
\bibitem[\protect\citeauthoryear{Yadav \& Pahari}{2013}]{b49} Yadav, J. S., \& Pahari, M., 2013, under preparation
\bibitem[\protect\citeauthoryear{Zdziarski \& Gierlin\'ski}{2004}]{b45} Zdziarski, A. A. \& Gierlin\'ski, M. 2004, Prog. Theor. Phys. Suppl., 155, 99
\end{thebibliography}
\end{document}